\documentclass[11pt,a4paper]{article}
\pdfoutput=1
\usepackage{jheppub,bm,booktabs,multirow}
\usepackage{xcolor}
\usepackage{overpic}
\usepackage{enumitem} 
\usepackage{verbatim}

\usepackage{graphicx}
\usepackage{subcaption}

\usepackage{adjustbox}
\usepackage{tikz}
\usetikzlibrary{decorations.pathreplacing}
\usetikzlibrary{arrows, arrows.meta}

\allowdisplaybreaks

\makeatletter
\def\@fpheader{~}
\makeatother

\usepackage[Q=yes,pverb-linebreak=no]{examplep}

\def\e{\epsilon}

\def\bT{\boldsymbol{T}}

\newcommand{\jets}{\mathrm{jets}}
\newcommand{\plist}{\{\underline{p}\}}

\newcommand{\nlist}{\{\underline{n}\}}
\newcommand{\zlist}{\{\underline{z}\}}
\newcommand{\F}{\mathcal{F}}

\newcommand{\bH}{\boldsymbol{\mathcal{H}}}

\newcommand{\bGam}{\boldsymbol{\Gamma}}
\newcommand{\bU}{\boldsymbol{U}}
\newcommand{\bP}{\mathbf{P}}

\newcommand{\thetain}{\ensuremath{\Theta_{\mathrm{in}}}}

\newcommand{\bS}{\boldsymbol{\mathcal{S}}}

\newcommand{\Rm}[1]{\ensuremath{\boldsymbol{R}_{#1}}}
\newcommand{\Vm}[1]{\ensuremath{\boldsymbol{V}_{#1}}}
\newcommand{\vm}[1]{\ensuremath{\boldsymbol{v}_{#1}}}
\renewcommand{\rm}[1]{\ensuremath{\boldsymbol{r}_{#1}}}
\newcommand{\dm}[1]{\ensuremath{\boldsymbol{d}_{#1}}}
\newcommand{\Eout}{\ensuremath{E_\mathrm{out}}}
\newcommand{\Etot}{\ensuremath{E_\mathrm{tot}}}
\newcommand{\ptot}{\ensuremath{\vec{p}_\mathrm{tot}}}
\newcommand{\ceps}{\tilde{c}^\epsilon}

\title{Factorization and Resummation for Sequential Recombination Jet Cross Sections}
\author[a]{Thomas Becher}
\author[b]{and J\"urg Haag}

\affiliation[a]{Albert Einstein Center for Fundamental Physics, Institut f\"ur Theoretische Physik, Universit\"at Bern, Sidlerstrasse 5, CH-3012 Bern, Switzerland}
\affiliation[b]{Physik Institut, Universit\"at Z\"urich, CH-8057 Z\"urich, Switzerland}

\emailAdd{becher@itp.unibe.ch}
\emailAdd{juerg.haag@physik.uzh.ch}

\date{\today}

\date{September 28, 2023}

\preprint{\begin{flushright}
ZU-TH 61/23\\ 
September 28, 2023
\end{flushright}}

\abstract{We extend the class of factorization theorems for non-global observables from fixed angular constraints to cross sections defined in terms of sequential jet clustering. The associated hard and soft functions depend not only on the directions of the hard partons, but also on their energy fractions. We derive the one-loop anomalous dimension of the hard functions that drives the leading-logarithmic resummation. The anomalous dimension imposes energy ordering, which simplifies the clustering sequence. We perform resummations for gap-between-jet observables defined with different jet algorithms and explain the effects of the clustering on the importance of secondary emissions and on the effective gap size.}

\begin{document}

\maketitle

\section{Introduction}

There has been significant progress in the theoretical description of non-global observables in the past few years. These observables involve angular cuts on soft radiation, which induces a complicated pattern of logarithmically enhanced higher-order corrections due to secondary emissions off hard partons. The paper by Dasgupta and Salam who 
discovered the effect more than twenty years ago also introduced a Monte Carlo (MC) method to compute the leading logarithms numerically in the large-$N_c$ limit \cite{Dasgupta:2001sh}. Over the past two years, several groups extended the resummation to subleading logarithmic accuracy \cite{Banfi:2021xzn,Banfi:2021owj,Becher:2021urs,Becher:2023vrh,FerrarioRavasio:2023kyg}. In addition, there are now also methods to carry out resummations accounting for the full color structure \cite{Weigert:2003mm,Hatta:2013iba,Hagiwara:2015bia,Hatta:2020wre,Platzer:2013fha,AngelesMartinez:2018cfz,Forshaw:2019ver}. One particularly interesting effect which is absent in the large-$N_c$ limit are Glauber phases. They destroy coherence and lead to the so-called super-leading logarithms \cite{Forshaw:2006fk}, for which first resummations are now available \cite{Becher:2023mtx,Boer:2023jsy}.

The resummations of super-leading logarithms and subleading logarithms in \cite{Becher:2021urs,Becher:2023vrh} are based on factorization theorems \cite{Becher:2015hka,Becher:2016mmh} derived in soft-collinear effective theory \cite{Bauer:2001yt,Bauer:2002nz,Beneke:2002ph}. The factorization theorems provide a framework to use renormalization group (RG) methods to carry out the resummation. While these methods have been applied to a variety of different observables \cite{Balsiger:2018ezi,Balsiger:2019tne,Chien:2019gyf,Balsiger:2020ogy,Becher:2022rhu}, they only apply to cross sections with fixed angular constraints, such as isolation cones or hemisphere jet masses. 

Most measurements at the large hadron collider (LHC) involve hadronic jets, typically defined using sequential clustering jet algorithms of the $k_t$-variety, i.e., $k_t$ \cite{Catani:1993hr,Ellis:1993tq}, anti-$k_t$ \cite{Cacciari:2008gp} or Cambridge/Aachen (C/A) \cite{Dokshitzer:1997in,Wobisch:1998wt} jets. The associated angular constraints differ order-by-order and the factorization theorems \cite{Becher:2015hka,Becher:2016mmh} do not apply. The goal of the present paper is to extend the factorization to recombination jet algorithms and to take into account clustering effects in the resummation. 

Such clustering effects were studied shortly after NGLs were discovered \cite{Appleby:2002ke} and it was found that they reduce the size of the non-global effects. However, not only the secondary but also primary emissions are affected by clustering as stressed in \cite{Delenda:2006nf}. In addition to gaps between jets, clustering effects were also investigated for $Z$+jet \cite{Delenda:2012mm} and $H$+jet \cite{Ziani:2021dxr}. Within the context of SCET, clustering effects were discussed in \cite{Kelley:2012kj,Kelley:2012zs}. Based on the fact that even in the abelian case the clustering constraints change order by order, these papers concluded that a resummation within the usual factorization into a product of hard $H$, jet $J$ and soft function $S$ was not possible. This is true, but even in the absence of clustering effects, factorization theorems of non-global observables do not take a simple product form, but involve an infinite tower of hard and soft functions \cite{Becher:2015hka,Becher:2016mmh}.

In our paper, we will show that it is relatively straightforward to generalize the factorization \cite{Becher:2015hka,Becher:2016mmh} to sequential recombination jet algorithms. For simplicity, we will focus on jets with large jet radius $R$, so that only soft radiation is relevant. Since the clustering is in general sensitive to the energy fractions of the partons, the hard and soft functions need to be computed differentially in the energy fractions of the hard partons in order to be able to formulate the clustering constraints. These are in general complicated, but significantly simplify when restricting the logarithmic accuracy. A major simplification arises because the one-loop anomalous dimension for leading logarithmic (LL) accuracy imposes strong energy ordering.\footnote{We consider single-logarithmic observables, so that LL refers to leading single logarithms.} Thus, at this accuracy, the only information affecting the clustering of the emissions is the order in which they are emitted and their direction. The first clustering effects with non-trivial  dependence on the energy ratios of emissions arise in the double emission part of the two-loop anomalous dimension~\cite{Becher:2021urs}, which is relevant for next-to-leading logarithmic (NLL) accuracy.

The factorization theorem will be derived in Section~\ref{sec:fact} after we review $k_t$ clustering algorithms in Section~\ref{sec:jetalgorithms}. The leading-order RG evolution generates strongly ordered emissions and we discuss in Section~\ref{sec:stronglyordered} how the jet clustering simplifies in this limit and can be added to existing numerical resummation codes. In Section~\ref{sec:gapsize} we then investigate the clustering effects in detail. We find that there are mainly two effects. First, the clustering increases the jet size by clustering soft radiation into the jet. Secondly, it reduces the effect of secondary radiation by clustering small-angle radiation. We conclude in Section~\ref{sec:conclusion}.

\section{Sequential recombination jet cross sections \label{sec:jetalgorithms}}

Before discussing the factorization theorem, we will define the class of observables under consideration, set up our notation, and review the jet algorithms we will focus on. We will consider a cross section with (at least) $M$ hard jets and will put a veto that restricts the radiation inside a certain angular region outside the jets, which will be referred to as the out-region or the gap. This type of observable is known as a gap-between-jets cross section. 

The focus of our paper is on jet clustering and to discuss its effects in the simplest possible setting, we will consider $e^+e^{-}$ cross sections, but our considerations also apply to the hadron collider case. The fixed-order cross section in  dimensional regularization with $d=4-2\epsilon$ takes the form 
\begin{align}\label{eq:sigma}
	\sigma(Q_0) = \frac{1}{2Q^2}\!\!\sum_{m=M}^\infty \prod_{i=1}^m \!\int \! [dp_i] |\mathcal{M}_m(\plist)|^2 (2\pi)^d\delta(Q-\Etot)\delta^{(d-1)}(\ptot) \F(\plist)\Theta(Q_0 - 2\Eout)\hspace{0.9pt} ,
\end{align}
where $Q$ is the center of mass energy, $M$ is the minimal number of hard jets and partons (e.g., $M=2$ for dijet observables), $Q_0$ is the veto energy, and $[dp_i]$ denotes the phase-space measure for parton $i$.
Below, we will use the color-space formalism and denote the $m$-parton amplitude by $|\mathcal{M}_m(\plist)\rangle$, where $\plist = \{p_1, \dots, p_m\}$ so that
\begin{equation}
|\mathcal{M}_m(\plist)|^2= \langle\mathcal{M}_m(\plist)|\mathcal{M}_m(\plist)\rangle\,.
\end{equation}
The cross section involves two types of constraints. The function
$\F(\plist)$ is an arbitrary infrared safe measurement function, depending on hard scales of order $Q$. For example, if we wanted to calculate the cross section differential in the energy of the leading jet $E_\mathrm{jet,1}$, we would use $\F(\plist)=\delta(y-E_\mathrm{jet,1}(\plist))$.
The cross section $\sigma(Q_0)$ in \eqref{eq:sigma} will depend on the hard scales in the function $\F(\plist)$, but we will keep this dependence implicit. For the derivation of the factorization theorem in Section~\ref{sec:fact}, we will assume that $\F$ does not involve large scale ratios.

To define $\Eout$, the energy in the out-region, we apply a clustering algorithm on the partons $\{p_1,\dots p_m\}$, which returns a list of jets $\{P_1,\dots,P_{n_{\jets}} \}$. We then define
\begin{align}\label{eq:Eout}
    \Eout = \sum_{j=1}^{n_{\jets}} P_j^0\, \Theta_{\mathrm{out}}(P_j)\, ,
\end{align}
where $\Theta_{\mathrm{out}}(P_j)$ is zero, if the jet is inside and one if it is outside. 
Alternatively, one could define the observable $\Eout$ as the energy of the leading  jet in the gap. The LL resummation is insensitive to the exact definition. A dependence on the precise definition of $\Eout$ only arises at higher logarithmic accuracy, through corrections to the soft functions. 

Two common definitions of the out-region are the following: i) One can consider all additional jets (or particles that are not part of the $M$ hardest jets) to be out and restrict their energy. We stress that all exclusive jet cross sections involve a veto on additional jets, so this type of constraint is quite common. ii) One can define a certain angular region outside the $M$ hard jets and restrict the energy there. The angular region will be defined in terms of the directions of the hard jets, e.g., by defining the gap to be a certain angular distance away from all hard jets. Below, we will consider both possibilities i) and ii). In both cases, the out condition depends on the momenta of the hard jets. For brevity, we do not indicate this in \eqref{eq:Eout}, but the reader should keep it in mind.


For the jet clustering algorithms we will concentrate on the final state version of the inclusive $k_t$-type clustering algorithms \cite{Catani:1993hr,Ellis:1993tq,Dokshitzer:1997in,Wobisch:1998wt,Cacciari:2008gp}, which we now briefly summarize:\footnote{See \cite{Salam:2010nqg} for a detailed review of different jet algorithms and their properties.} 
\begin{enumerate}
    \item For a list of partons with momenta $\{p_1,\dots,p_n\}$, determine the distances
    \begin{align}
        d_{ij}&=\min(E_i^{2p},E_j^{2p})\frac{\Delta R^2_{ij}}{R^2}\,, \quad i\neq j\in \{1,\dots,n\}\notag\\
        d_i&=E_i^{2p}\,, \quad i \in \{1,\dots,n\}\, ,
    \end{align}
    where $p=1$ is the $k_t$ algorithm, $p=0$ is the Cambridge-Achen (C/A) algorithm and $p=-1$ is the anti-$k_t$ algorithm. The angular distance $\Delta R^2_{ij}$ is a dimensionless symmetric function that vanishes if $i$ and $j$ become collinear. The parameter $R$ is referred to as clustering distance or jet radius. 
    \item Find the minimum of the $d_{ij}$ and $d_i$.
    \item If it is a $d_{ij}$, combine the two partons into a single one with combined momentum $p_{ij}=p_i+p_j$ and return to step 1. 
    \item Otherwise, if the minimum is a $d_i$, declare the corresponding particle to be a jet, remove it from the list of particles, and return to 1.
    \item Stop when no particles remain.
\end{enumerate}
The standard choice for the angular distance at lepton colliders is $\Delta R^2_{ij}=2(1-\cos \theta_{ij})$, where $\theta_{ij}$ is the angle between the two particles. At hadron colliders, one instead uses $\Delta R^2_{ij}=(y_i-y_j)^2+(\phi_i-\phi_j)^2$, where $y$ and $\phi$ denote the rapidity and azimuthal angle of the particles with respect to the beam axis. To work with the hadron-collider angular measure at an $e^+e^-$ collider, one can define the rapidity using the thrust axis. 

There are other possible recombination schemes instead of simply adding the momenta $p_{ij}=p_i+p_j$ in step three. Our framework can also accommodate other prescriptions, such as the winner-take-all scheme \cite{Bertolini:2013iqa,Larkoski:2014uqa}. 
    
A crucial observation is that the clustering algorithm commutes with an overall rescaling of the momenta. For example, if the jet clustering algorithm on a list of momenta $\{p_1,\dots,p_m\}$ returns the jets $\{P_1,\dots, P_{n_{\jets}}\}$, then running the same algorithm on the particles $\{\lambda p_1,\dots,\lambda p_m\}$ will return the jets $\{\lambda P_1,\dots, \lambda P_{n_{\jets}}\}$. This implies, that the clustering sequence only depends on the directions of the partons $\{p_1,\dots, p_n\}$ and the ratios of their energies.

\section{Factorization\label{sec:fact}}
The theoretical basis for the factorization theorem is the method of regions expansion for phase-space and loop integrals, together with the standard factorization of soft emissions from hard partons. For $Q_0\ll Q$, the partons in the gap are necessarily soft and we can expand the scattering amplitudes accordingly. The amplitude with $m$ hard partons and an arbitrary number of soft partons can be obtained by taking a matrix element of 
\begin{equation}\label{eq:softFact}
\bm{S}_1(n_1)\,\bm{S}_2(n_2)\dots \,\bm{S}_m(n_m) |{\cal M}_{m}(\{\underline{p}\})\rangle\,,
\end{equation}
where 
\begin{equation}\label{eq:Si}
\bm{S}_i(n_i) = {\bf P} \exp\left( ig_s \int_0^{\infty}\!ds\,
   n_i\cdot A_s^a( s n_i)\,\bm{T}_i^a \right)
\end{equation}
is a Wilson line along the direction $n_i^\mu = p_i^\mu/E_i$ of the hard parton in the appropriate color representation $\bm{T}_i^a$. The Wilson line $\bm{S}_i(n_i)$ is a matrix that acts on the color index of parton $i$.

The factorization theorems in \cite{Becher:2015hka,Becher:2016mmh} were obtained by inserting the representation \eqref{eq:softFact} into the cross section formula \eqref{eq:sigma} and expanding the phase-space constraints in the soft momenta. In doing so, the cross section splits into a hard function and a soft function defined as 
\begin{align}
    \bS_m(\nlist,\zlist, Q_0)=\int\limits_{X}\hspace{-0.55cm}\sum\bigl\langle 0\bigl|\bS_1^{\dagger}\left(n_1\right) \ldots \bS_m^{\dagger}\left(n_m\right)\bigr| X\bigr\rangle\bigl\langle X\bigl|\bS_1\left(n_1\right) \ldots \bS_m\left(n_m\right)\bigr| 0\bigr\rangle \theta\left(Q-2\Eout\right).
\end{align} 
The soft function contains Wilson lines for both the amplitude and its conjugate. 
In the papers \cite{Becher:2015hka,Becher:2016mmh} the out-region was fixed so that the soft function was only sensitive to the directions $\nlist$ of the $m$ hard partons. With sequential clustering, the situation is more complicated. As detailed in the previous section, the shape of the out-region depends nontrivially on the energy fractions $\zlist=\{z_1,z_2,\dots, z_m\}$ of the partons since these affect the clustering. 

To obtain a factorization theorem for sequential clustering jet cross sections, we therefore need to make the hard functions differential in the energy ratios of the colored partons.  We find it convenient to take the partons $\{p_1,\dots,p_m\}$ in the hard function to be ordered in energy with $E_1>E_2>\dots>E_m$. We then introduce the energy fractions as  $z_i=E_i/E_{i-1}\in [0,1]$ with $E_0=Q/2$. With this, we define the hard functions as
\begin{align}\label{eq:Hm}
\bH_m(\nlist,\zlist,Q)&=
\frac{1}{2Q^2}\Biggl(\prod_{i=1}^m \int \frac{d E_i E_i^{d-3}}{\ceps(2 \pi)^2}\Biggr) \widetilde{\bH}_m(\plist) \times \notag\\
&\quad
(2\pi)^d\delta(Q-\Etot)\delta^{d-1}(\ptot) \F(\plist)\thetain (\plist) \prod_{j=1}^m \delta\left(z_j-\frac{E_j}{E_{j-1}}\right)\,,
\end{align}
where the ``unintegrated'' hard function
\begin{equation}
 \widetilde{\bm{\mathcal{H}}}_m(\{\underline{p}\})  = |\mathcal{M}_m(\{\underline{p}\})\rangle \langle\mathcal{M}_m(\{\underline{p}\})| 
\end{equation}
is simply the squared amplitude summed/averaged over the particle spins. Up to the $\delta$-function constraints on the energy fractions $z_j$, the hard functions agree with the definitions in \cite{Becher:2015hka,Becher:2016mmh,Becher:2021urs}. The constant $\tilde{c} = e^{\gamma_E}/\pi$ was introduced into the hard functions in \cite{Becher:2021urs}. We suppress a sum over the permutations of the partons that we need because the momenta $\plist$ are ordered in energy. There are in general several hard functions of multiplicity $m$, therefore the sum over $m$ implicitly includes a sum over all partonic final states. We note that the renormalized hard function will be distribution valued in both the angles of the partons and the $\zlist$-variables.

In terms of the functions defined above, the factorization theorem takes the form
\begin{align}\label{eq:fact}
\sigma(Q, Q_0) &=  \sum_{m=M}^\infty \big\langle \bH_m(\nlist,\{\underline{z}\},Q,\mu) \otimes_z \bS_m(\nlist,\{\underline{z}\},Q_0,\mu) \big\rangle\, ,
\end{align}
 where the angular brackets denote the color trace $\langle M\rangle= \operatorname{tr}(M)$. The symbol $\otimes_z$ indicates that one has to integrate both over the energy fractions $\zlist$ and over the directions of the $m$ hard partons in $\bH_m(\nlist,\zlist, Q)$, which are the same as the directions of the Wilson lines in $\bS_m(\nlist, \zlist , Q_0)$, i.e.,\ 
\begin{align}
    \bH_m(\nlist,\zlist, Q) \otimes_z & \,\bS_m(\nlist,\zlist, Q_0)=\\
    &  \prod_{i=1}^m \int\!\left[d \Omega_i\right] \int_0^1 \!dz_i \,\bH_m(\nlist,\zlist, Q)\, \bS_m(\nlist,\zlist, Q_0)\, ,
\end{align}
where we use the shorthand notation 
\begin{equation}
    \int\left[d \Omega_i\right]=\tilde{c}^\epsilon \int \frac{d^{d-2} \Omega_i}{2(2 \pi)^{d-3}}\, .
\end{equation}
Note that the factor $\tilde{c}^\e$ cancels between the angular integral and the energy integration and together one recovers the usual phase-space integral in $d$ dimensions. For the present paper, we focus on the case $R\sim 1$. For small jet radii, one can factorize the hard and soft functions further to resum logarithms of $R$.

The main physics reason, why a factorization theorem into soft and hard physics still holds, lies in the fact that the clustering of the hard partons is not disturbed by the presence of additional soft partons. To be precise, if a set of hard partons $\plist$ are clustered into the hard jets $\{P_1\dots,P_{n_{\jets}}\}$, then additional soft partons $\{q_1,\dots,q_{n_{\mathrm{S}}}\}$ will, at leading power in the soft expansion,  never change the hard jet momenta, even if they cluster together with hard partons. Those soft partons which do not cluster with the hard ones will form soft jets $\{Q_1\dots,Q_{n_{S,\jets}}\}$ and we will end up with a final list of jets $\{P_1\dots,P_{n_{\jets}},Q_1\dots,Q_{n_{S,\jets}}\}$, where the momenta of the hard jets remain unchanged by the soft radiation. In particular, we will not suddenly end up with a hard jet in the out region.

\section{RG evolution and resummation\label{sec:resummation}}

In our effective-theory framework, the hard functions $\bH_m$ are the Wilson coefficients of the operator matrix elements $\bS_m$. The operator matrix elements suffer from ultraviolet divergences, which are absorbed into the Wilson coefficients and render them finite, but dependent on the renormalization scale $\mu$.

Independence of the cross section of the renormalization scale then leads to the RG equations
\begin{align}\label{eq:RG}
\frac{d}{d \ln \mu} \bH_m(\nlist,\zlist, Q, \mu) & =-\sum_{l=M}^m \bH_l(\nlist,\zlist, Q, \mu)\, \bGam_{l m}^H(\nlist,\zlist, Q, \mu)\,,
\end{align}
which form the basis of the resummation of large logarithms of $Q/Q_0$. For the case of fixed-cone cross sections, the derivation of $\bGam^H$ was discussed in great detail in \cite{Becher:2021urs}. The anomalous dimension is obtained by considering soft limits of the hard amplitudes and extracting the associated singularities. 

When viewed as an infinite matrix in the space of particle multiplicities, the one-loop anomalous dimension for dijet production, $M=2$, takes the form
\begin{equation}
    \bGam^H 
    =\frac{\alpha_s}{4 \pi} \left(\begin{array}{ccccc}
\Vm{2} & \Rm{2} & 0 & 0 & \ldots \\
0 & \Vm{3} & \Rm{3} & 0 & \cdots \\
0 & 0 & \Vm{4} & \Rm{4} & \cdots \\
0 & 0 & 0 & \Vm{5} & \cdots \\
\vdots & \vdots & \vdots & \vdots & \ddots
\end{array}\right)+\ldots.
\end{equation}
The entries $\Vm{m}$ and $\Rm{m}$ correspond to the soft singularities in virtual and real emissions. The real emission entries increase the number of hard partons by one, i.e., $\Rm{m}$ maps a hard function with $m$ partons onto one with $(m+1)$ hard partons. The anomalous dimension matrix at the two-loop level contains three different entries $\vm{m}$, $\rm{m}$ and $\dm{m}$, associated with double-virtual, real-virtual and double-real soft singularities \cite{Becher:2021urs}. The mixing of different parton multiplicities makes the solution of \eqref{eq:RG} highly non-trivial.

Let us now extract the one-loop anomalous dimension. The determination of the virtual terms is of course unaffected by the energy fraction constraint in \eqref{eq:Hm} and the result reads \cite{Becher:2016mmh,Becher:2021urs}
\begin{align}\label{eq:Vm}
    \Vm{m}=2 \sum_{(i j)}\left(\bT_{i, L} \cdot \bT_{j, L}+\bT_{i, R} \cdot \bT_{j, R}\right) \int\left[d \Omega_q\right] W_{i j}^q\, ,
\end{align}
where we used that the imaginary Glauber phases cancel for $e^+e^-$-colliders. The color generator $\bT^a_{i, L}$ acts on the amplitude, while $\bT^a_{i, R}$ acts on its conjugate, i.e., on the right-hand side of the hard function. Through the dipole radiator
\begin{align}
\label{eq:Wijk}
    W_{ij}^q=\frac{n_i\cdot n_j}{n_i \cdot n_q n_j \cdot n_q}\, 
\end{align}
the virtual part \eqref{eq:Vm} of the anomalous dimension depends on the directions $\nlist$ of the emitting dipoles, but it is independent of their energy fractions $\zlist$. 

The energy fractions come into play for the real emission part $\Rm{m}$ obtained by analyzing the hard function with $(m+1)$ partons in the limit where one of the partons becomes soft. Since we ordered the hard function by energy, the one becoming soft is the last parton whose momentum we will denote by $q\equiv p_{m+1}$. The hard function reads
\begin{align}
& \mathcal{H}_{m+1}\left(\left\{\underline{n}, n_{q}\right\},\left\{\underline{z}, z_{q}\right\}, Q, \epsilon\right)=\frac{1}{2 Q^2} \prod_{i=1}^{m+1} \int \frac{d E_i E_i^{d-3}}{\tilde{c}^\epsilon(2 \pi)^2}\delta(z_m-\frac{E_i}{E_{i-1}}) \nonumber\\
&\hspace{1.8cm}  \times\widetilde{\bm{\mathcal{H}}}_{m+1}(\{\underline{p},q\})(2 \pi)^d \delta\big(Q-\sum_{i=1}^{m+1} E_i\big) \delta^{(d-1)}\left(\vec{p}_{\text {tot }}+\vec{q}\right) \thetain(\{\underline{p}, q\})
\end{align}

Since we are only interested in the soft singularity arising from $E_{m+1} \to 0$, we can put an upper cutoff $\Lambda \ll Q$ on the corresponding energy integral and then expand the integrand in the small energy $E_q \equiv E_{m+1}$. The amplitude factorizes as
\begin{equation}
\widetilde{\bm{\mathcal{H}}}_{m+1}(\{\underline{p},q\}) = \frac{1}{E_{q}^2} \, W_{ij}^q \, \bT_i^a \, \widetilde{\bm{\mathcal{H}}}_{m}(\{\underline{p}\}) \,\bT_i^{\tilde{a}} \,,
\end{equation}
where $a$ and $\tilde{a}$ are the color indices of the emitted gluon in the amplitude and its conjugate. After expanding the small momentum out of the phase-space constraints and approximating $\delta(z_{q}-{E_{q}}/{E_m})$ by $\delta(z_{q})$, we can carry out the integration over $E_q$ and find that the soft divergence takes the form
\begin{equation}
\begin{aligned}
\label{eq:Hmp1NLOIR}
&\bH_{m+1}\left(\left\{\underline{n}, n_q\right\},\left\{\underline{z}, z_{q}\right\}, Q\right)  \\
&\hspace*{2cm}= \delta(z_{q})\frac{2}{\epsilon} \frac{\alpha_s}{4 \pi} \thetain\left(n_q\right) \sum_{(i j)} W_{i j}^q \, \bT_i^a \,\bH_{m+1}\left(\left\{\underline{n}\right\},\left\{\underline{z}\right\}, Q\right) \,\bT_j^{\tilde{a}}\, .
\end{aligned}
\end{equation}
To obtain this result, it was crucial that the angular constraint in the hard function factorized as $\thetain(\{\underline{p}, q\})=\thetain(\{\underline{p}\})\thetain(n_q)$, which is trivial for the fixed angular regions considered in \cite{Becher:2016mmh,Becher:2021urs}. In our case, the $\thetain(n_q)$ condition for the softest parton will depend on the directions and energy fractions of all harder partons through the clustering, but it is important that the constraint on the hard partons $\thetain(\{\underline{p}\})$ is unchanged by the presence of the soft parton. This is the case because the clustering of soft partons does not change the directions of the hard jets, as we have explained at the end of Section~\ref{sec:fact}.

From the coefficient of the $1/\e$-pole in  \eqref{eq:Hmp1NLOIR}, we can immediately read off the anomalous dimension 
\begin{align}\label{eq:Rm}
    \Rm{m}=-4\, \delta(z_{q})\sum_{(i j)} \bT_{i, L}^ a \bT_{j, R}^{\tilde{a}} W_{i j}^q \thetain\left(n_q\right)\, .
\end{align}
Except for the presence of the factor $\delta(z_{q})$, this anomalous dimension looks identical to the one in \cite{Becher:2016mmh,Becher:2021urs}, but we stress that $\thetain\left(n_q\right)$ not only depends on the direction of $q$ but has a complicated dependence on $\zlist$ and $\nlist$ of the $m$ hard partons. 

There is an issue with the extraction of the anomalous dimension that was swept under the rug in the above discussion, namely that the individual hard functions also contain collinear singularities. Indeed, the dipole $W_{ij}^q$ in the results for \eqref{eq:Rm} and \eqref{eq:Vm} becomes singular when $q$ becomes collinear to the direction of one of the emitting partons $i$ or $j$. For $e^+e^-$ colliders, the collinear contributions cancel out in the cross section \cite{Becher:2021urs}, while the initial-state collinear singularities give rise to SLLs in hadronic collisions. The treatment of the collinear singularities was discussed in detail in \cite{Becher:2023mtx}. For the present paper, we make use of the fact that the cancellation takes place and assume that the collinear singularities are separately regularized at an intermediate stage of the computation. This is indeed what is done in the parton shower solution of \eqref{eq:RG}, where the collinear singularities of $\Vm{m}$ and $\Rm{m}$ are regularized with an angular cutoff.

With the anomalous dimension at hand, we can now resum large logarithms. To do so, one computes the hard functions $H_l$ at a scale $\mu_h\sim Q$, where they are free of large logarithms. Then one solves the RG to evolve the functions down to a scale $\mu_s \sim Q_0$, where the soft functions are evaluated. The resummed cross section is obtained as
\begin{align}
\sigma(Q_0)=\sum_{l=M}^{\infty}\Bigl\langle\bH_l\left(\nlist,\zlist, Q, \mu_h\right) \sum_{m \geq l} \bU_{l m}\left(\nlist,\zlist,\mu_s,\mu_h\right) \otimes_z \bS_m\left(\nlist,\zlist, Q_0, \mu_s\right) \Bigr\rangle \,,
\end{align}
where the integrations in $\otimes_z$ involve the directions and energy fractions of all $m$ partons in the soft function.
Since the hard and soft functions are free of large logarithms, one can expand them in $\alpha_s$. At the lowest order, all soft functions are trivial $\bS_m = \bm{1}$, and all hard functions $\bH_l$ with $l>M$ are suppressed. The evolution matrix takes the form
\begin{align}
    \bU\left(\nlist,\zlist,  \mu_s, \mu_h\right)=\bP \exp \left[\int_{\mu_s}^{\mu_h} \frac{d \mu}{\mu}\bGam^H(\nlist,\zlist, \mu)\right] .
\end{align}
The path-ordering symbol $\mathbf{P}$ is necessary since $\bGam^H$ is a matrix. For the resummation of the leading single logarithms, one evolves with the one-loop matrix, and it is convenient to introduce the ``shower time'' $t\equiv t(\mu_h,\mu_s)$ through
\begin{equation}
\int_{\mu_s}^{\mu_h} \frac{d\mu}{\mu}\, \bm{\Gamma}^H= \int_{\alpha(\mu_s)}^{\alpha(\mu_h)} \frac{d\alpha}{\beta(\alpha)}\, \frac{\alpha}{4\pi}\,\bm{\Gamma}^{(1)} =\frac{1}{2\beta_0}\ln\frac{\alpha(\mu_s)}{\alpha(\mu_h)}\,\bm{\Gamma}^{(1)} = t\,\bm{\Gamma}^{(1)}  \,.
\end{equation}
and the shorthand notation
\begin{align}
\bm{\mathcal{H}}_m(t) \equiv \bm{\mathcal{H}}_M(\nlist,\zlist,Q,\mu_h)\, \bm{U}_{Mm}(\nlist,\zlist,\mu_h,\mu_s)\,
\end{align}
for the leading-logarithmic hard functions obtained after RG evolution.

In terms of the shower time $t$, the RG equation for these functions reads
\begin{align}\label{eq:diffhrd}
\frac{d}{dt}\,\bm{\mathcal{H}}_m(t)  &=   \bm{\mathcal{H}}_m(t) \,  \bm{V}_m +   \bm{\mathcal{H}}_{m-1}(t) \,  \bm{R}_{m-1} \, ,
\end{align}
which has the iterative solution \cite{Balsiger:2018ezi}
\begin{align}\label{eq:iterRG}
\bm{\mathcal{H}}_M(t) &= \bm{\mathcal{H}}_M(0) \,e^{t \bm{V}_M} \,,\nonumber\\
\bm{\mathcal{H}}_{M+1}(t) &= \int_{0}^{t} dt' \,\bm{\mathcal{H}}_{M}(t') \, \bm{R}_{M}\, e^{(t-t')  \bm{V}_{M+1}}\,, \\
\bm{\mathcal{H}}_{M+2}(t) &= \int_{0}^{t} dt' \,\bm{\mathcal{H}}_{M+1}(t') \, \bm{R}_{M+1}\, e^{(t-t')  \bm{V}_{M+2}}\,, \nonumber \\
\bm{\mathcal{H}}_{M+3}(t) &= \dots \,. \nonumber 
\end{align}
The leading-logarithmic cross section is then obtained by summing the different multiplicities and integrating over the directions. For dijet production $M=2$ we have
\begin{align}\label{eq:sigmaLL}
\sigma_{\mathrm{LL}}(t) &= \sum_{m=2}^\infty \big\langle  \bm{\mathcal{H}}_m(t) \,\otimes_z\, \bm{1} \big\rangle\nonumber \\
&= \big\langle \bm{\mathcal{H}}_2(t) + \int \!dz_{3} \int \! \left[d \Omega_{3}\right] \bm{\mathcal{H}}_{3}(t) \nonumber \\ 
&\hspace{1.7cm} +  \int \! dz_{3}  \int \!\left[d \Omega_{3}\right]\int \! dz_{4} \int \! \left[d \Omega_{4}\right]   \bm{\mathcal{H}}_{4}(t) + \dots \big\rangle \,,
\end{align}
where we have explicitly written out the integrations over the angles and energy fractions of the additional emissions generated by the shower. The integrals over the directions and energy fractions of the original $M$ partons are constrained by momentum conservation and the phase-space constraints in $\F(\plist)$. For dijet production, $M=2$, the original partons will be back-to-back and carry half of the center-of-mass energy, therefore their energy fractions will be $z_1=z_2=1$. For convenience, we have normalized the hard functions to the Born-level cross section 
\begin{equation}
\sigma_{\mathrm{B}} = \langle \bm{\mathcal{H}}_2(0) \rangle\,.
\end{equation}
In addition to the cross section, it is also interesting to consider the gap fraction
\begin{equation}
\Sigma(t) = \sigma_{\mathrm{LL}}(t)/ \sigma_{\mathrm{B}}\,,
\end{equation}
which measures the effect of the veto on the cross section. By definition $\Sigma(0)=1$.

Since the anomalous dimension $\Rm{m}$ in \eqref{eq:Rm} involves a $\delta$-function in the energy fraction of the emitted parton, the $z$-integrations in \eqref{eq:sigmaLL} can immediately be carried out. The only information which remains is that successive emissions are always infinitely softer. After performing the $z$-integrations, \eqref{eq:iterRG} and \eqref{eq:sigmaLL} are well suited for MC evaluation, at least if one resorts to the large-$N_c$ limit where the color structure is trivial. The corresponding MC evolves in $t$ and iteratively generates additional partons along randomly chosen directions, thereby MC sampling the angular integrals in \eqref{eq:sigmaLL}. Since the $z$-integrals are trivial, the shower is the same as in the fixed-cone case, and the detailed steps to reformulate all individual factors in \eqref{eq:iterRG} and \eqref{eq:sigmaLL} can be found in \cite{Balsiger:2018ezi,Balsiger:2020ogy}.  At each time step, the shower keeps a list of directions, ordered according to their large-$N_c$ dipole structure. The only additional information needed in our case is the energy ordering of the emissions, which is needed as an input to perform the clustering. In the next section, we discuss the clustering in the strongly ordered limit in detail for the different jet algorithms and explain how it can be efficiently implemented in the shower. 

At LL level, nontrivial dependence on the energy fraction arises through the hard function for $M\geq 3$ jets, however, the integrals are simply the ones present in the Born level phase-space. The associated energy dependence can be taken into account, for example by using a tree-level event generator for the $M$-jet cross section and then showering the individual events, as was done e.g.\ in \cite{Balsiger:2018ezi,Balsiger:2020ogy}.

Nontrivial $z$-integrals in the shower only arise at NLL accuracy, through the two-loop anomalous dimension, specifically in the part $\bm{d}_m$, which describes double real emissions. These emissions are unordered and the associated contribution will involve an integral over the relative energy fraction. The integral over this fraction will be non-trivial, even for C/A jets, because the direction of the jet depends on the energies in the case where the two unordered particles cluster. The purely virtual contribution encoded in $\bm{v}_m$ is of course unaffected by the clustering and the real-virtual terms $\bm{r}_m$ will be proportional to $\delta(z_q)$, as the one-loop real-emission result \eqref{eq:Rm}.

\section{Strongly ordered clustering\label{sec:stronglyordered}}
To perform the LL resummation for jet-clustering algorithms, we need to understand how the $k_t$-type algorithms simplify in the limit of strongly ordered soft emissions. We already explained in Section~\ref{sec:fact}, that adding a much softer parton $p_{m+1}$ to a list of partons $\{p_1,\dots,p_m\}$ does not change the jet momenta $\{P_1,\dots, P_{n_J}\}$ one obtains from clustering the original partons at leading power in the soft expansion. The new momentum $p_{m+1}$ can either be clustered with one of the harder partons, in which case its momentum can be neglected, or it forms its own jet. This observation is very useful when we use a Monte Carlo shower to perform the resummation. The shower works by generating more and more strongly ordered emissions until an emission does not satisfy the in-constraint, in which case the shower is stopped. This allows us to analyze in- and out-conditions recursively. In particular, if a new emission $p_{m+1}$ gets clustered with any other parton, we immediately know that the in-condition is satisfied. If $p_{m+1}$ becomes a jet, we need to check a simple constraint for this jet only. For instance, for our setup i), any new jet would immediately count as out, while for our setup ii), one would only need to check whether the new jet falls into the fixed angular gap.

To give more details on how the strongly ordered clustering can be implemented in the shower, we now discuss the three $k_t$-type algorithms separately. We will always start with a list of partons $\{p_1,\dots,p_m\}$ that satisfy the in-condition, and we add a yet much softer parton $p_{m+1}$ for which we need to decide whether it satisfies the in-condition. We explained in Section~\ref{sec:resummation} that the $z$-integrals are trivial at LL. Thus, we only carry around the directions $n_i$ of the parton $p_i$ in the shower. We will often refer to the parton by its direction in what follows.
\subsection{Anti-$k_t$ algorithm}
Since the anti-$k_t$-distances
\begin{align}
        d_{ij}&=\min(E_i^{-2},E_j^{-2})\frac{\Delta R^2_{ij}}{R^2}\, ,\notag\\
        d_i&=E_i^{-2}
\end{align}
involve negative powers of the energies, the first $m$ partons will prefer to first cluster amongst themselves before they cluster with $p_{m+1}$. This means, that we can replace the first $m$ parton momenta with the jets $\{P_1,\dots,P_{M},\dots\}$ with directions $\{N_1,\dots,N_{M},\dots\}$  that one gets from clustering them.\footnote{Depending on the in-condition, there may or may not be more than the original $M$ hard jets. In particular, for scenario i), where only the $M$ hardest jets count as in, no extra jets are possible for in-configurations.} Then, we can check if the extra parton clusters with one of the hard jets. This is the case if and only if there exists a jet $P_j$ such that $\Delta R^2(P_j,p_{m+1})=\Delta R^2(N_j,n_{m+1})\leq 1$. Geometrically, this means that the region where the direction $n_{m+1}$ is clustered with a harder parton is given by the union of circles with radius $R$ around the jet directions $\{N_1,\dots,N_{M},\dots\}$. We visualize this clustering condition in Figure~\ref{fig:Clust_Akt}. In the shower, the clustering can be efficiently implemented by keeping a list of the jet directions $\{N_1,\dots,N_{M},\dots\}$, which are present at a given shower step. 

\subsection{$k_t$-type algorithm}
Since the $k_t$-distances
\begin{align}
        d_{ij}&=\min(E_i^{2},E_j^{2})\frac{\Delta R^2_{ij}}{R^2}\,,\notag\\
        d_i&=E_i^{2} 
\end{align}
involve positive powers of the energies, the minimal distance in the first step of the clustering algorithm will involve the softest parton $p_{m+1}$. It will cluster with one of the harder $m$ partons if there exists a parton momentum $p_j$ satisfying $\Delta R^2(p_j,p_{m+1})=\Delta R^2(n_j,n_{m+1})<1$, else $p_{m+1}$ becomes a jet itself. Geometrically, this means that the region where the direction $n_{m+1}$ is clustered with a harder parton is given by the union of circles with radius $R$ around the parton directions $\{n_1,\dots,n_{m}\}$. We visualize this clustering condition in Figure~\ref{fig:Clust_kt}. In the shower, this condition is straightforward to implement, as we need to carry around all parton directions anyway. 
\subsection{C/A algorithm}
\begin{figure}
    \centering
    \begin{subfigure}{0.3\textwidth}
        \includegraphics[width=\textwidth]{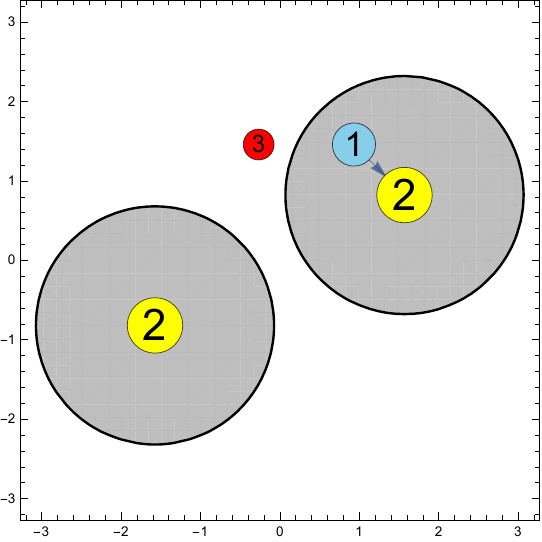}
        \caption{ Anti-$k_t$}
        \label{fig:Clust_Akt}
    \end{subfigure}
    \begin{subfigure}{0.3\textwidth}
        \includegraphics[width=\textwidth]{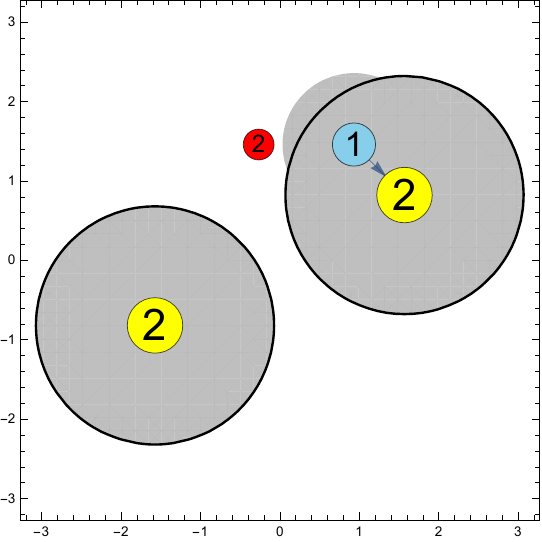}
        \caption{C/A}
        \label{fig:Clust_CA}
    \end{subfigure}
    \begin{subfigure}{0.3\textwidth}
        \includegraphics[width=\textwidth]{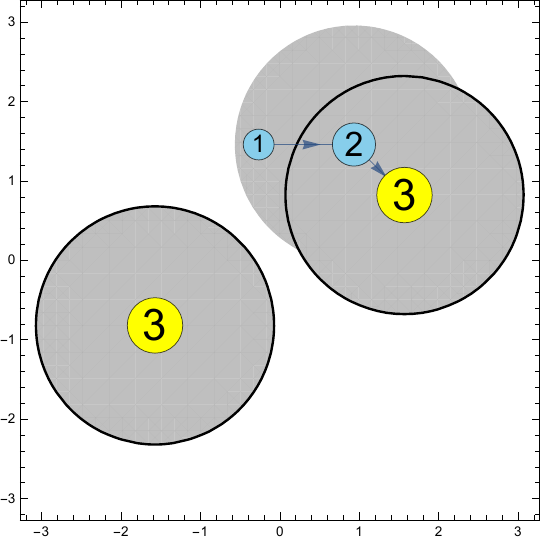}
        \caption{$k_t$}
        \label{fig:Clust_kt}
    \end{subfigure}
    \caption{
    Clustering of a four-particle configuration  arising in the parton shower performing the resummation for dijet production. The $x$- and $y$-axes correspond to azimuth and rapidity of the partons, respectively, and we cluster according to $\Delta R_{ij}^2=\Delta\eta^2+\Delta \phi^2$.  The shower starts with two back-to-back hard jets, shown as yellow disks and generates two strongly-ordered soft emissions. The energy ordering of the emissions is indicated by their disk size. For the given configuration, the first emission always gets clustered with the primary jets. The second emission forms a new jet for the anti-$k_t$ and the C/A algorithms, while it is clustered with the primary jets for the $k_t$ algorithm (blue disk). The numbers in the disks indicate at which step the corresponding particle gets clustered with a harder particle (in which case an arrow indicates the harder particle), or becomes a jet. If the numbers are the same, then the clustering distances are the same and the order of the steps is ambiguous. For $k_t$ clustering, the softest emission is clustered first, while it is clustered last for anti-$k_t$. In gray, we show the effective area of the jets, i.e., the region where the second emission would be clustered with the hard jets. The area is largest for $k_t$ and smallest for anti-$k_t$, for which it agrees with fixed cone clustering.}
    \label{fig:clust}
\end{figure}

The C/A algorithm is the most complicated one out of the three. The reason is, that the C/A-distances
\begin{align}
    d_{ij}&=\frac{\Delta R^2_{ij}}{R^2}\,,\notag\\
    d_i&=1
\end{align}
do not produce a strong order amongst any of the parton distances. Let us first understand, why the two strategies used for the $k_t$ and anti-$k_t$ algorithms would break down for the C/A algorithm. If we would simply combine $p_{m+1}$ with another parton $p_j$ if it is closer than $R$, we would sometimes mistakenly cluster the two. Such a mistake would happen if $p_j$ was clustered with an even harder parton at an earlier stage in the clustering (c.f. Figure \ref{fig:Clust_CA}). If, on the other hand, we would first cluster the $\{p_1,\dots,p_{m}\}$ into jets and only then cluster $p_{m+1}$ with those jets, we would sometimes mistakenly classify $p_{m+1}$ as a jet (c.f. the gray area in Figure \ref{fig:Clust_CA}).

An efficient algorithm for the shower, that takes care of these subtleties, works as follows: Together with the list of parton directions $\{n_1,\dots,n_m\}$, we also carry around a list of distances $\{\underline{\delta}\}=\{\delta_1,\dots,\delta_m\}$, where there is one $\delta_i$ for each $n_i$. The $\delta_i$ have the following meaning: If the corresponding parton with direction $n_i$ is never clustered with a harder parton (i.e., if it is also a jet), we set it to 1. In particular, all the original hard partons $\{n_1,\dots,n_{M}\}$ are also jets, and their $\delta_i$ are thus to be set to 1. If, on the other hand, $n_i$ gets clustered with a harder momentum, say $n_j$ we set $\delta_i=d_{ij}$. To decide whether a new direction $n_{m+1}$ is in or out, we notice that $n_{m+1}$ can only cluster with $n_i$, if $d_{i(m+1)}<\delta_i$ as otherwise $n_i$ would cluster with a harder parton before it can cluster with $n_{m+1}$. To be more precise, $n_{m+1}$ will cluster with the parton $n_i$ that is closest to $n_{m+1}$, while also satisfying $d_{i(m+1)}<\delta_i$. Geometrically, this means that the region where the direction $n_{m+1}$ is clustered with a harder parton is given by the union of circles with radii $\{\delta_1,\dots,\delta_{m}\}$ around the parton directions $\{n_1,\dots,n_{m}\}$. We visualize this clustering condition in Figure~\ref{fig:Clust_CA}. If we find that $n_{m+1}$ clusters with $n_i$, we add it to the list of parton directions and we also add the distance $d_{i(m+1)}$ to $\{\underline{\delta}\}$. If there is no $n_i$ that can cluster with $n_{m+1}$, $n_{m+1}$ becomes a jet. If the jet is out, the shower is stopped. Else, we again add $n_{m+1}$ to the list of parton directions, and we add $d_{m+1}=1$ to $\{\underline{\delta}\}$.

The above clustering prescriptions are easily implemented into the parton shower framework used for the resummation and are much more efficient than performing the full clustering algorithm again after every emission. Using geometric information, the asymptotic efficiency could be improved further at the cost of a more complicated implementation. This might be useful to study the asymptotic behavior at very large shower times $t$ but is not necessary for phenomenologically accessible values of $t$.

\section{Clustering effect and effective gap-size\label{sec:gapsize}}

In the following, we will analyze the impact of jet clustering effects for the energy flow into a gap between dijets in $e^+e^-$ collisions in detail. We will study both types of gap observables introduced in Section~\ref{sec:jetalgorithms}, the case i), where the gap is defined as the entire region outside the two hard jets, and ii) where the gap is a fixed angular region outside the two hard jets. All our numerical results are obtained in the strict leading color limit ($N_c\to\infty$ at fixed $\alpha_s N_c$).

\begin{figure}
    \centering
    \includegraphics[width=0.7\textwidth]{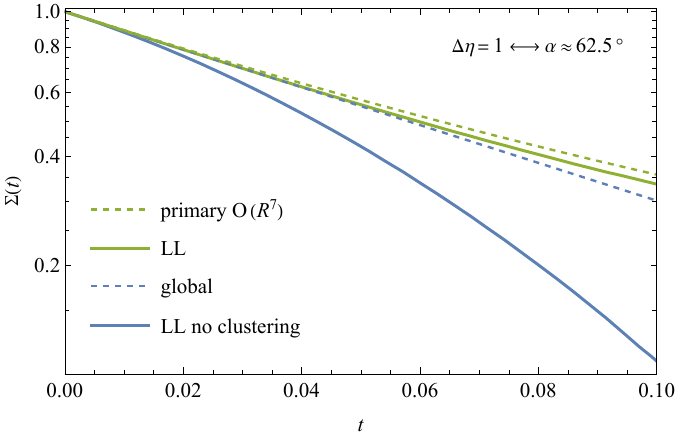}
    \caption{
    Effect of $k_t$ jet-clustering on the gap fraction for a fixed central rapidity gap of size $\Delta\eta=1$. 
    }
    \label{fig:Delenda-Comparison}
\end{figure}

As a warm-up, we follow \cite{Delenda:2006nf} and consider the case ii) with a fixed central rapidity gap of size $\Delta\eta=1$, where the rapidity $\eta$ and the azimuth $\phi$ of an emission are defined with respect to the thrust axis. We perform $k_t$ clustering using the angular distance $\Delta R^2_{ij}=(y_i-y_j)^2+(\phi_i-\phi_j)^2$. A jet counts as out if it ends up in the rapidity gap. This set-up was analyzed in \cite{Delenda:2006nf} because it allows for analytic resummation of primary emission clustering logarithms, which was carried out to $O(R^7)$. We show our results for this set-up in Figure~\ref{fig:Delenda-Comparison}. From top to bottom we plot the result for the analytically resummed primary emissions derived in \cite{Delenda:2006nf} (up to $O(R^7$)), the full LL result computed with our Monte Carlo shower code, the naive exponentiation of the one-emission result (labelled ``global'' in the plot), and the $R=0$ limit of the full LL contribution, which is the LL result for the gap fraction without applying any clustering.

Our results perfectly agree with the ones shown in \cite{Delenda:2006nf}.\footnote{Note that our definition of $t$ differs by a factor of $1/2$ from the one in \cite{Delenda:2006nf}} One sees that the clustering algorithm drastically reduces the suppression of the gap fraction. The fact that the full LL result is close to the result obtained by taking into account only primary emissions (dashed green line) was interpreted in \cite{Delenda:2006nf} as an indication that the non-global logarithms are strongly suppressed by the clustering. Interestingly, clustering effects increase the gap fraction compared to what one would expect from a naive exponentiation of the one-emission result (dashed blue line), which we will call the global contribution in the following. At phenomenologically relevant values of $t\lesssim 0.07$ corresponding to $Q_0\gtrsim 1 \,\mathrm{GeV}$ for a hard scale of $Q= 100\,\mathrm{GeV}$, the relative difference between the full LL and the primary emission gap fraction is approximately a $5\%$ effect. For this choice of parameters, the naive exponentiation is closer to the full result than including the primary emissions with clustering. However, this is just a coincidence specific to the choice of parameters. 

Using a distance measure with rapidity defined with respect to the thrust axis simplified the analytical computations in \cite{Delenda:2006nf}, but implies that the angular distance $\Delta R_{ij}^2$ between the primary two partons $n_1$ and $n_2$ and any other emission is infinite (at least at LL), so that emissions can never cluster with the primary two jets. To study the phenomenologically more relevant scenario i), we now switch to the standard lepton-collider angular distance  $\Delta R_{ij}^2=2 n_i\cdot n_j=2(1-\cos\theta_{ij})$, where $\theta_{ij}$ is the angle between the directions $n_i$ and $n_j$ in the lab-frame. With this distance measure, the hardest two partons are not special in terms of angular separation and can cluster with other partons. In fact, with the gap defined as the outside of the two hardest jets, for emissions to be counted as in, they have to be clustered together with $n_1$ or $n_2$ at one point.

\begin{figure}
  \centering
\begin{subfigure}[b]{\textwidth}
    \centering
    \includegraphics[width=0.65\textwidth]{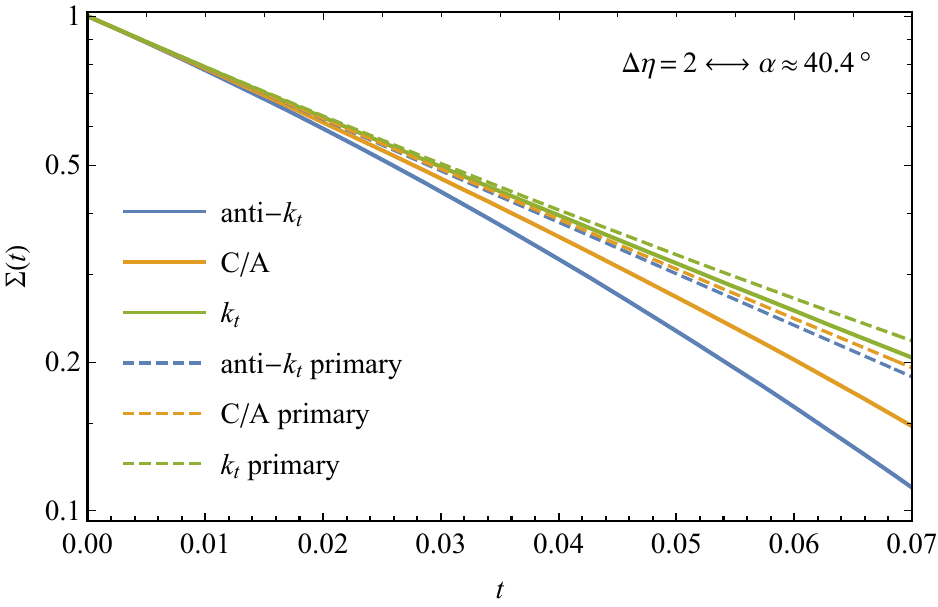}
  \vspace*{0.2cm}\end{subfigure}
  
\begin{subfigure}[b]{\textwidth}
    \centering
    \includegraphics[width=0.65\textwidth]{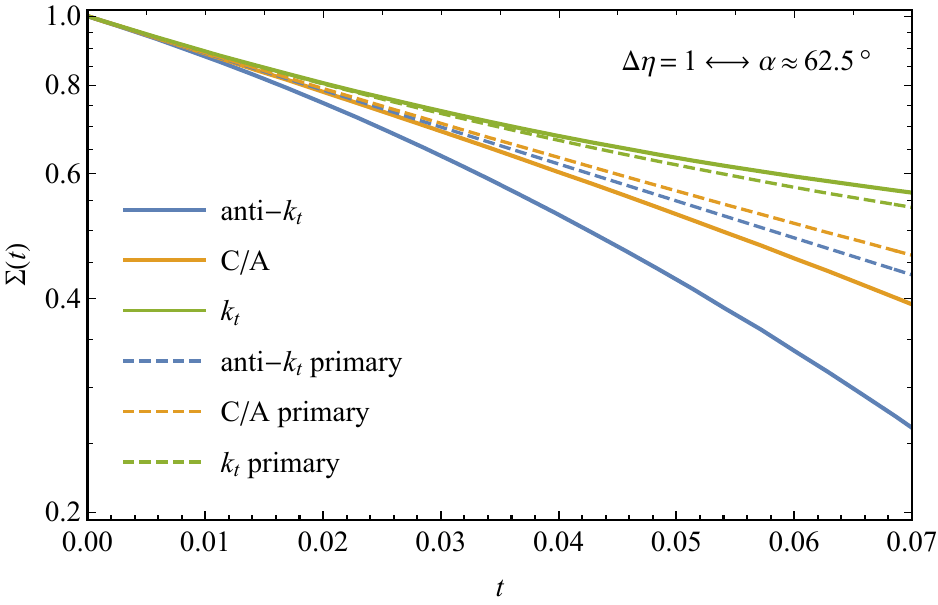}
  \vspace*{0.2cm}\end{subfigure}  
   
\begin{subfigure}[b]{\textwidth}
    \centering
    \includegraphics[width=0.65\textwidth]{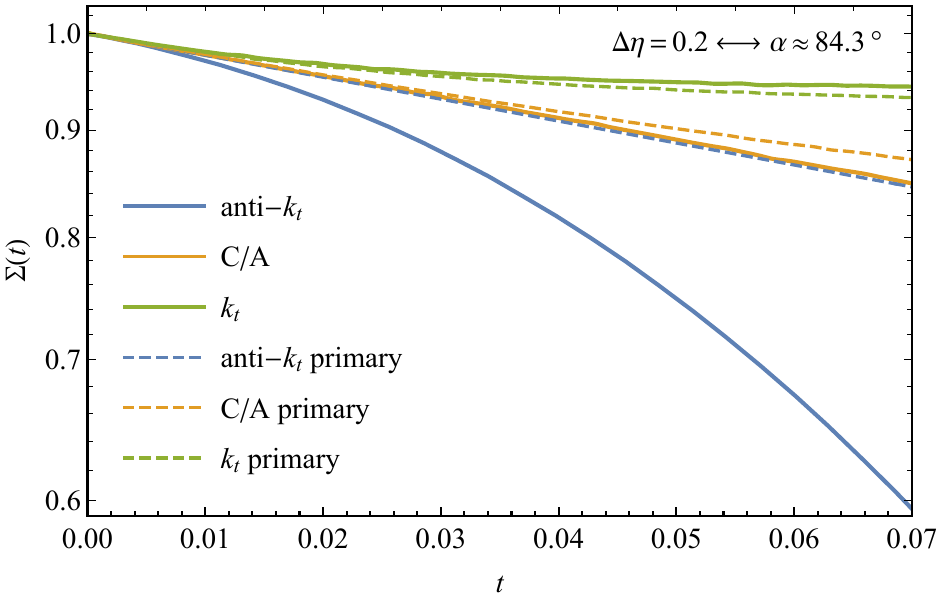}
  \end{subfigure}

  \caption{Jet clustering effects for different gap sizes in dijet production. We consider three different gap sizes with corresponding jet radius $R$ and jet opening angle $\alpha$. In addition to the shower results, we also give results if only emissions from the primary $q\bar{q}$-pair are included.}
  \label{fig:cos_gap_fractions}
\end{figure}

For this setup, the LL resummation for the anti-$k_t$ algorithms is particularly simple: soft emissions are in if and only if they fall into cones with opening half-angle $\alpha=\cos^{-1}(1-\frac{R^2}{2})$ around the primary back-to-back jets $n_1$ and $n_2$ and thus, at LL, the anti-$k_t$ algorithm behaves just as if we had put fixed cones around the thrust axis. Equivalently, the anti-$k_t$ algorithm with clustering distance $R$ behaves as if we had not clustered at all and just put a central rapidity gap 
\begin{equation}\label{eq:eta_vs_alpha_vsR}
    \Delta \eta = \ln\!\left(\frac{1+\cos\alpha}{1-\cos\alpha}\right)=\ln\!\left(\frac{4-R^2}{R^2}\right)
\end{equation}
with respect to the thrust axis. In Figure~\ref{fig:cos_gap_fractions} we show the gap fractions with this setup for the $k_t$-type clustering algorithms for three different choices of the jet radius $R$. For each clustering algorithm, we have also plotted the corresponding approximate gap fraction one obtains by only considering emissions from the primary partons. Due to the absence of clustering effects, the primary contribution is the same as the global result for the anti-$k_t$ algorithm. For all three choices of the clustering parameter, we find that the anti-$k_t$ algorithm leads to the smallest gap fraction, while the $k_t$ algorithm leads to the largest one, and the C/A algorithm lies somewhere in the middle. Compared to the naive global exponentiation, for the $k_t$ algorithm, the enhancement of the gap fraction due to clustering effects beats the suppression due to non-global effects for all three gap sizes we consider. For the C/A algorithm, the clustering enhancement only wins for very large clustering distances. We also observe that the primary contribution approximates the full result better for the $k_t$ algorithm than for the C/A algorithm, and in general, neglecting secondary radiation works best for large clustering distances $R$. Last but not least, we point out that for large $R$, the Sudakov suppression of the gap fraction becomes very weak. This might not be very surprising at first glance, because even for pure non-global effects, i.e., in the anti-$k_t$ case, the gap fraction $\Sigma(t)$ becomes smaller for small gap sizes. However, in the absence of clustering effects, the cross section always tends to $0$ for large $t$. With clustering the situation is different, in the last panel of Figure~\ref{fig:cos_gap_fractions} it seems that the gap fraction for the $k_t$ algorithm approaches a constant. It is not clear whether the same will be true for the C/A gap fraction at large $t$. In the following, we will try to understand the physics that explains these observations.

\subsection{Effective gap area}
We first want to understand why Sudakov effects on the gap fraction are suppressed so strongly for the $k_t$ and C/A algorithms and why $\Sigma(t)$ seems to approach a constant for asymptotically large times, at least for the $k_t$ algorithm. To understand the underlying reason, it is instructive to consider the limit of an infinitesimally small gap. In this case, virtual and real emissions cancel exactly, and the gap function is $1$ for all values of $t$. Physically speaking, this is obvious, as we are not imposing any veto in this situation. In terms of the parton shower, this can also be understood. Up to a normalization, the gap fraction $\Sigma(t)$ can be interpreted as the total probability to not emit a gluon into the gap before the shower time $t$. If there is no gap, then the probability not to emit into the gap is always a 100\%. 

Let us now try to analyze the behavior of the gap fraction in the last panel of Figure~\ref{fig:cos_gap_fractions}, in particular, let us focus on the $k_t$-algorithm. At shower time $t=0$, we start with an event $E=\{n_1,n_2\}$, where only the original back-to-back hard directions are present. Letting the shower evolve, we will at some time $t_1$ emit a gluon with direction $n_3$. The gluon will be in if $\min(d_{13},d_{23})<1$, which for the present choice of $R$ is equivalent to not emitting into a central rapidity gap of size $\Delta\eta=0.2$. Since the gap is so small, it is quite unlikely that we emit into the gap, nevertheless, it is possible and for small $dt$, one can calculate $\Sigma(dt)=1-4N_c\Delta\, \eta\, dt + O(dt^2)$, which is simply the expansion of the global approximation. If we continue running the shower from the time $t_1$, assuming that the emission $n_3$ is in, we will at some time $t_2$ emit a second gluon with direction $n_4$. We again need to decide what is the probability to emit into the gap. However, now the in-condition is $\min(d_{14},d_{24},d_{34})<1$. In other words, the effective gap is much smaller. In fact, for very small initial gap sizes $\Delta \eta$, one can convince oneself, that the effective gap is essentially half as large as the original gap wherever $n_3$ is, unless $n_3$ is very collinear to the primary directions, in which case one cannot distinguish $n_3$ from a virtual radiation. This implies, that after a first emission happened, the probability to emit into the gap in a small time-step $dt$ is essentially half of what it was before the first emission. For subsequent emissions, the effective gap size further decreases until we arrive at a shower time, say $t_n$, where the gap effectively vanishes. The probability that we do not generate any further emissions into the gap after $t_n$ is $1$. This mechanism explains the behavior of the gap fractions for the $k_t$ algorithm plotted in the last panel of Figure~\ref{fig:cos_gap_fractions}. With increasing $t$, the gap fraction initially decreases, until we arrive at a value where the effective gap essentially vanishes and the gap fraction plateaus. For the C/A algorithm, the behavior is similar, however since the out-condition is slightly weaker for C/A, i.e., some events that are in for $k_t$ are still out for C/A, the effective gap size decreases more slowly. One would maybe expect, that the gap fraction also plateaus for the C/A algorithm for large values of $t$. However, we did not find any evidence for this. A possible explanation is given in Section~\ref{sec:Suppression}.
\begin{figure}
    \centering
    \begin{subfigure}{0.49\linewidth}
        \includegraphics[width=\linewidth]{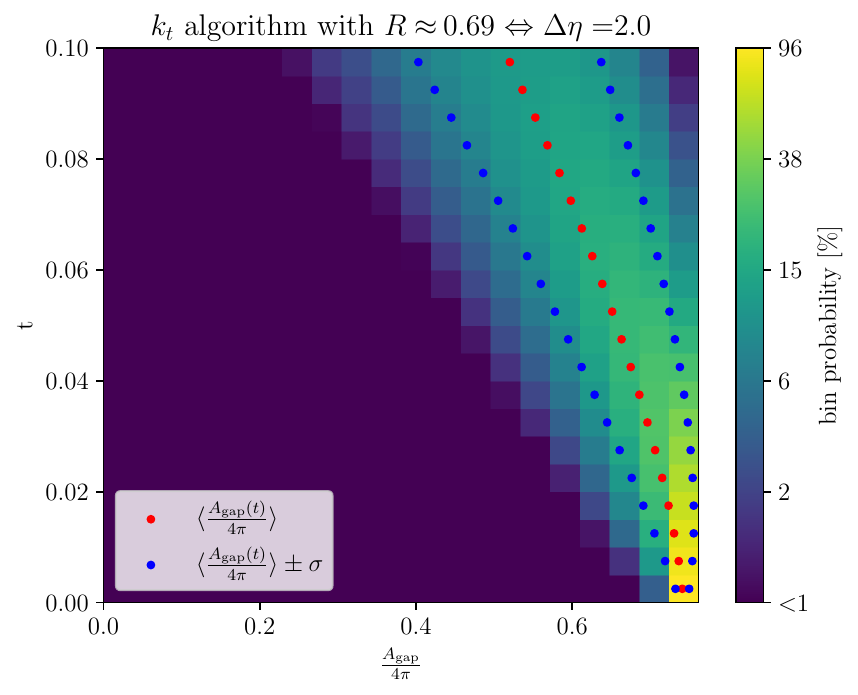}
    \end{subfigure}
    \hfill
    \begin{subfigure}{0.49\linewidth}
        \includegraphics[width=\linewidth]{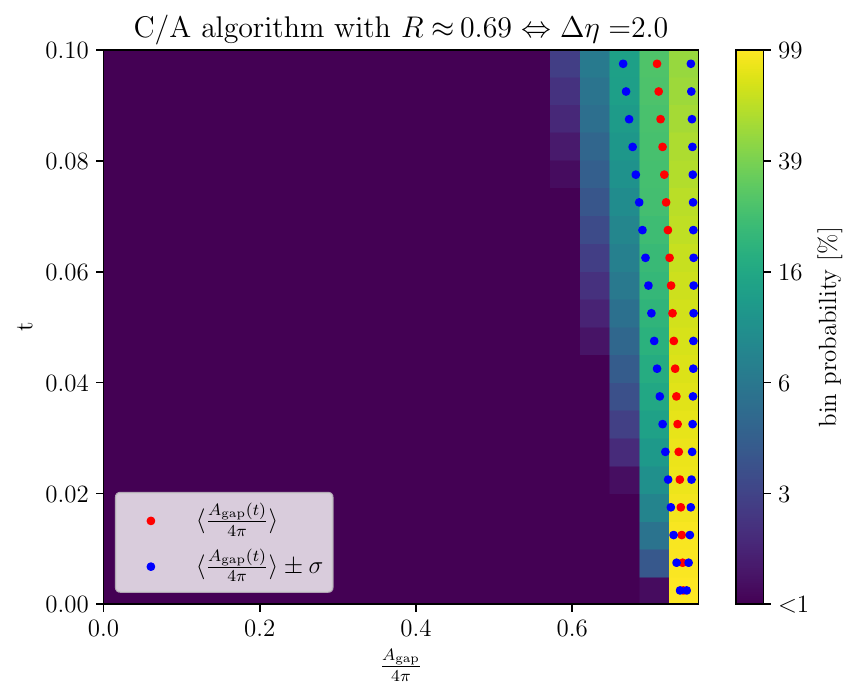}
    \end{subfigure}
    
    \medskip
    
    \begin{subfigure}{0.49\linewidth}
        \includegraphics[width=\linewidth]{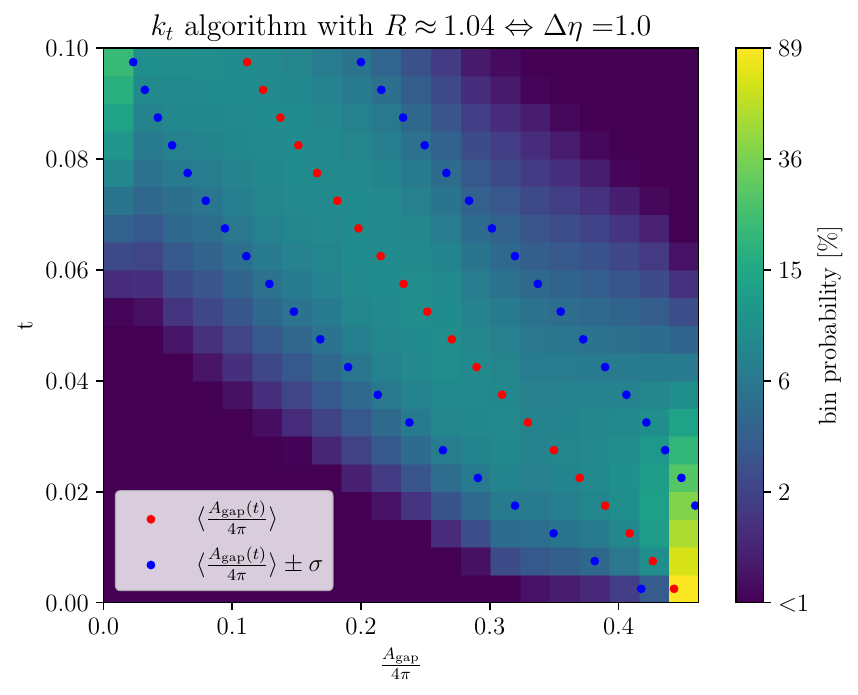}
    \end{subfigure}
    \hfill
    \begin{subfigure}{0.49\linewidth}
        \includegraphics[width=\linewidth]{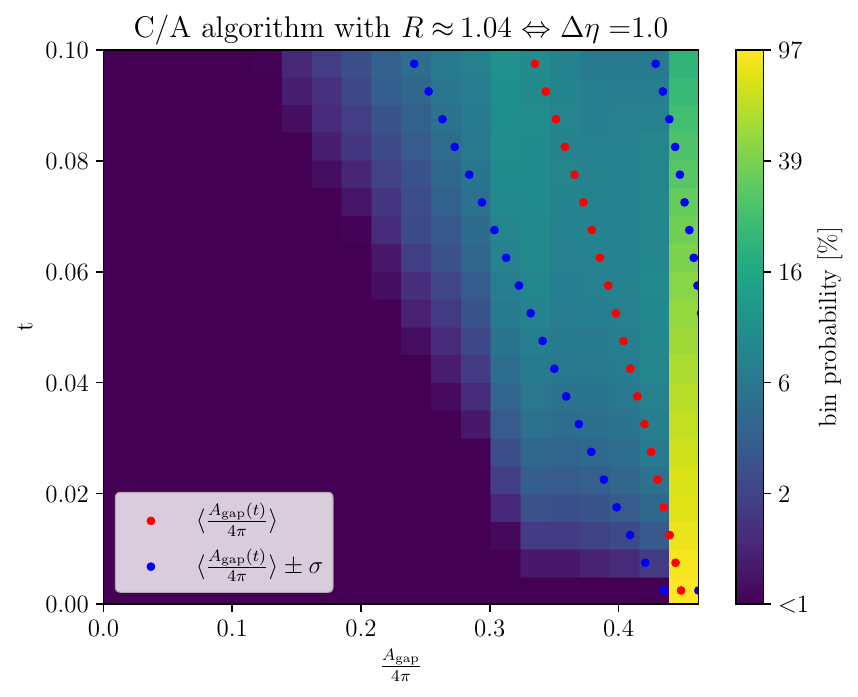}
    \end{subfigure}
    
    \medskip
    
    \begin{subfigure}{0.49\linewidth}
        \includegraphics[width=\linewidth]{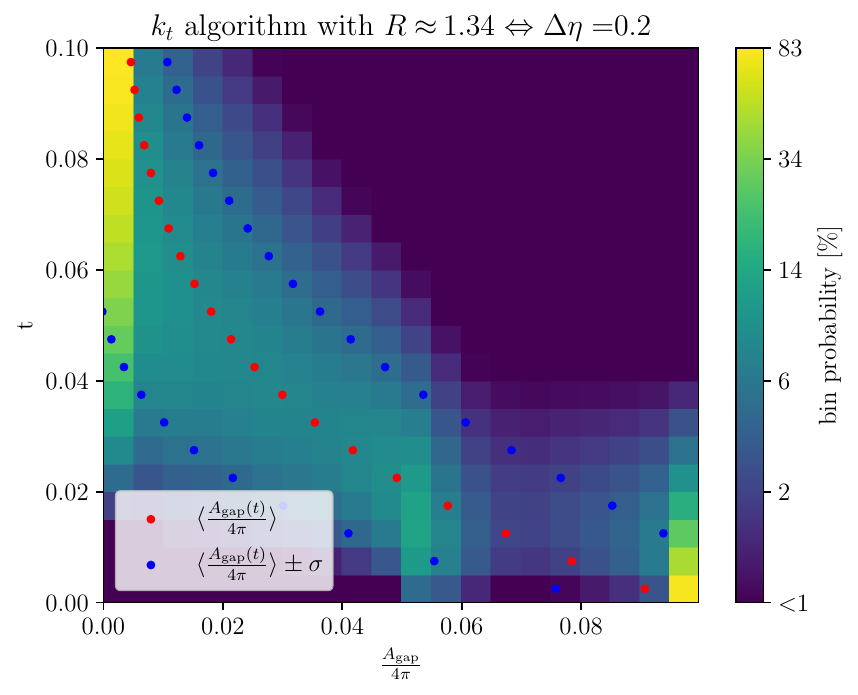}
    \end{subfigure}
    \hfill
    \begin{subfigure}{0.49\linewidth}
        \includegraphics[width=\linewidth]{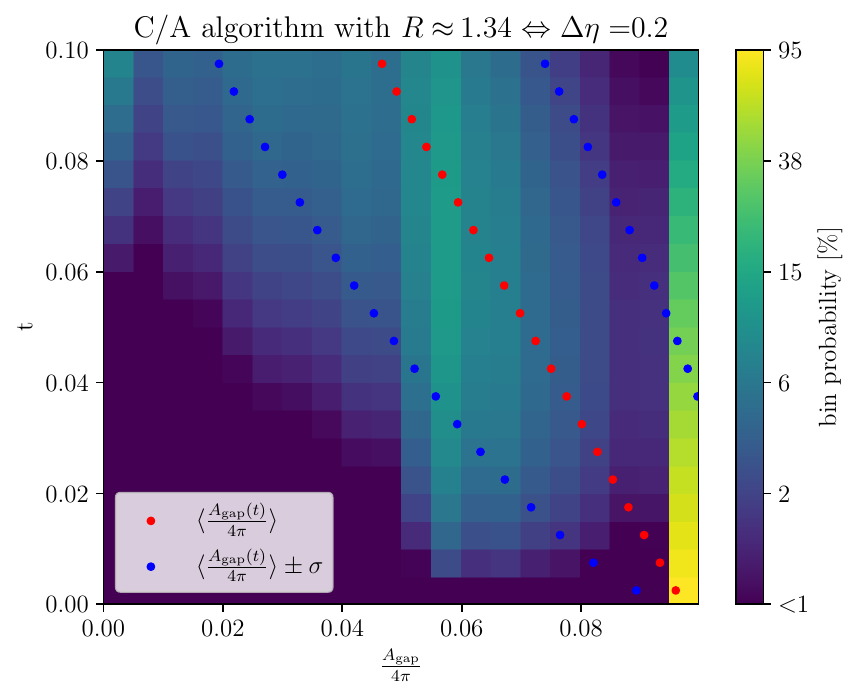}
    \end{subfigure}
    
    \caption{Probability distributions for the effective area of the out-region $A_{\mathrm{gap}}$ at shower time $t$ for different choices of the clustering algorithm and clustering distance $R$.}
    \label{fig:effective_area_plots}
\end{figure}

The shrinking of the out-region can be analyzed with the parton shower as follows. Before starting the shower, we set up a double differential histogram in the shower time $t$ and the out-area $A_{\mathrm{gap}}(t)$, and we fill the histogram over many shower events. Whenever the shower generates a new emission with direction $n_i$, say at shower time $t$, we calculate the area $A_{\mathrm{gap}}(t)$ on the sphere where $n_i$ would end up in the out-region, and we add the event to the corresponding bin of the histogram. After the shower finishes, we normalize each row of the histogram, corresponding to some shower time $t$, by the total number of events in that row.\footnote{The total number of events in a row corresponds to $\Sigma(t)$ up to normalization.} In Figure~\ref{fig:effective_area_plots} we show what one obtains for this histogram for different setups. A single subplot corresponds to a choice of the clustering algorithm and a choice of the clustering distance $R$. We chose the value of $R$ such that the out-region for the first emission corresponds to a central rapidity gap $\Delta\eta$ of 0.2, 1 or 2. The plots have to be read as follows: Each horizontal slice corresponds to a fixed shower time $t$. The areas on the sphere are calculated with the standard measure $d\cos\theta d\phi$. The effective area at time $0$ is given by 
\begin{align}
    A_{\mathrm{gap}}(0)=4\pi\cos\alpha=4\pi\left(1-\frac{R^2}{2}\right)=4\pi\ln\!\Big(\frac{e^{\Delta \eta}-1}{e^{\Delta \eta}+1}\Big)\, .
\end{align}
 For each shower time $t$, the color represents the probability, that the available effective area for out-emissions falls into one of 20 equal-sized bins from $0$ to $A_{\mathrm{gap}}(0)$. The $x$-axis has been normalized by the total area of the unit sphere $4\pi$. We also plot the mean value and the standard deviation of the effective area as a function of the shower time in red and blue, respectively. The shrinking of the effective area in shower time is clearly visible, and it is particularly fast for the $k_t$-algorithm.

\subsection{Suppression of secondary emission effects}
\label{sec:Suppression}
We have now understood, why Sudakov effects weaken for clustering algorithms when we get to large $t$ and small gap sizes. However, the shrinking of the effective gap size does not explain, why non-global effects due to secondary emissions are suppressed so strongly and why including only primary emissions seems to work so well for $k_t$ and C/A clustering. We summarize in Table~\ref{tab:primary_approximation} how well the primary contribution approximates the full result for our second set-up at shower time $t=0.07$. We see, that for the anti-$k_T$ algorithm, which at LL is equivalent to not clustering at all and using fixed cones, the primary (in this case global) approximation massively overshoots the correct result. For instance, for the choice $\Delta\eta=2.0$, we overshoot the correct LL result by more than 60\%. This result is of course well understood -- while the global contribution to the gap fraction is $\Sigma_\mathrm{global}(t)=e^{-4N_c \Delta \eta \,t}$, the non-global contribution asymptotically falls off roughly as $\Sigma_{\mathrm{LL}}(t)\sim \Sigma_\mathrm{global}(t)\, e^{-C t^2}$, where $C$ a positive constant, thus the relative difference between the global and LL result becomes large at late $t$. Before we try to understand why the same does not happen for the other clustering algorithms, we want to understand the physical mechanism that leads to the strong suppression of the gap fraction at large $t$ due to non-global effects. Such a mechanism was already analyzed for instance in \cite{Dasgupta:2002bw}. We will not reiterate the explanation from \cite{Dasgupta:2002bw}, but our discussion is closely related in spirit. The key point is that the non-global effects are driven by collinear radiation into the gap. In particular, this means that the dominant contribution comes from a region of phase space, where a gluon inside the jet is close to the gap boundary and emits a gluon into the gap at small angle. The behavior of $\Sigma_{\mathrm{LL}}(t)$ as a function of $t$ is again best understood by looking at the dynamics of the parton shower. At time $t=0$, only the primary partons are available. The shower then generates new emissions, which prefer to go collinear to the existing partons. With increasing $t$, the shower starts to populate the region inside the jet. After some time, it fills the region close to the jet boundary, allowing for the possibility of a collinear emission into the gap. After the region near the gap boundary becomes densely populated, it becomes very likely that new emissions are generated into the gap, at which point the shower is stopped. This explains the strong suppression of $\Sigma_\mathrm{LL}(t)$ at large $t$.

\begin{table}[]
    \centering
    \begin{tabular}{c|ccc}
    $\Delta \eta$  & anti-$k_t$[\%] & C/A[\%] & $k_t$ [\%]\\
            \hline
    $2.0$     & $66.6 \pm 1.5$ &  $32.0 \pm 1.4$ & $9.1  \pm 1.0$\\
    $1.0$     & $64.7 \pm 1.0$ &  $17.8 \pm 0.8$ & $-4.7 \pm 0.6$\\
    $0.2$     & $42.8 \pm 0.6$ &  $3.9  \pm 0.5$ & $-1.2 \pm 0.5$
    \end{tabular}
    \caption{Deviation of the primary-emission result from the full LL resummation for the anti-$k_t$ (i.e. no clustering), C/A and $k_t$ algorithms with $\Delta R_{ij}^2=2(1-\cos\theta_{ij})$. The table lists the ratio $(\Sigma_{\mathrm{primary}}(t)-\Sigma_{\mathrm{LL}}(t))/\Sigma_{\mathrm{LL}}(t)$ at $t=0.07$. The three rows correspond to different choices of the gap size $\Delta \eta$, or equivalently, the clustering distance $R$, see \eqref{eq:eta_vs_alpha_vsR}.}
    \label{tab:primary_approximation}
\end{table}

\begin{figure}
    \centering
    \begin{subfigure}{0.32\linewidth}
        \includegraphics[width=\linewidth]{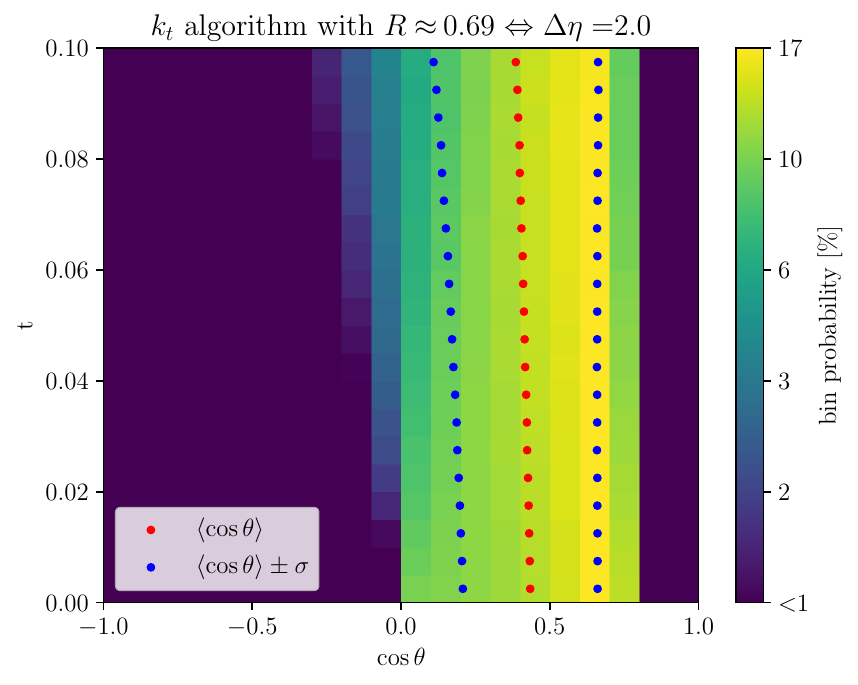}
    \end{subfigure}
    \hfill
    \begin{subfigure}{0.32\linewidth}
        \includegraphics[width=\linewidth]{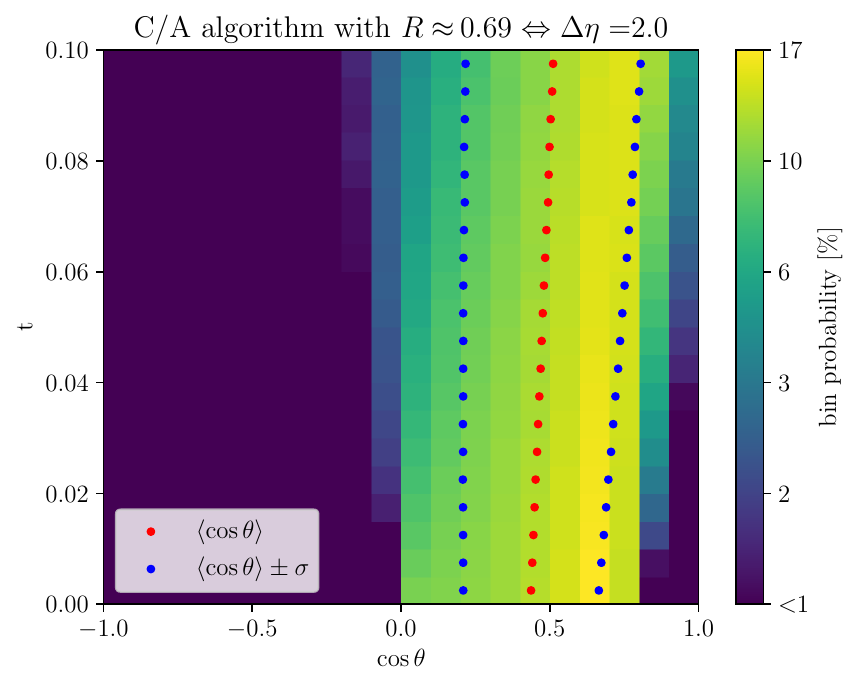}
    \end{subfigure}
    \hfill
    \begin{subfigure}{0.32\linewidth}
        \includegraphics[width=\linewidth]{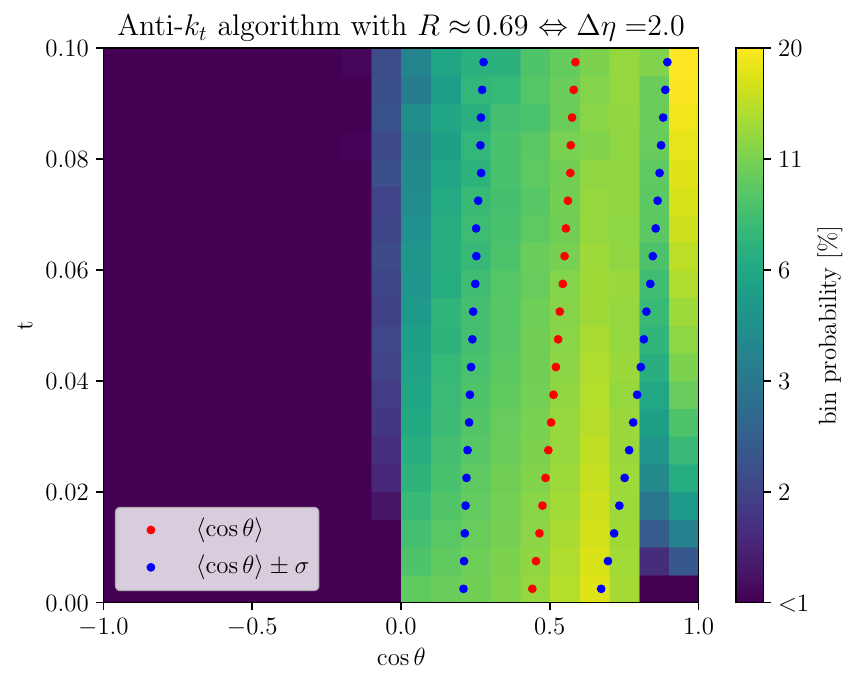}
    \end{subfigure}
    \medskip
    
   \begin{subfigure}{0.32\linewidth}
        \includegraphics[width=\linewidth]{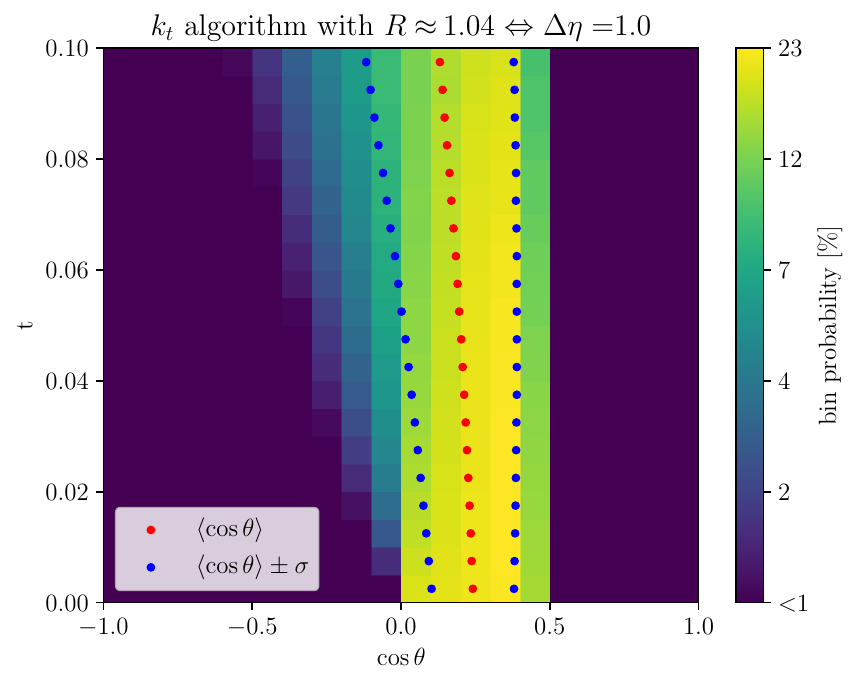}
    \end{subfigure}
    \hfill
    \begin{subfigure}{0.32\linewidth}
        \includegraphics[width=\linewidth]{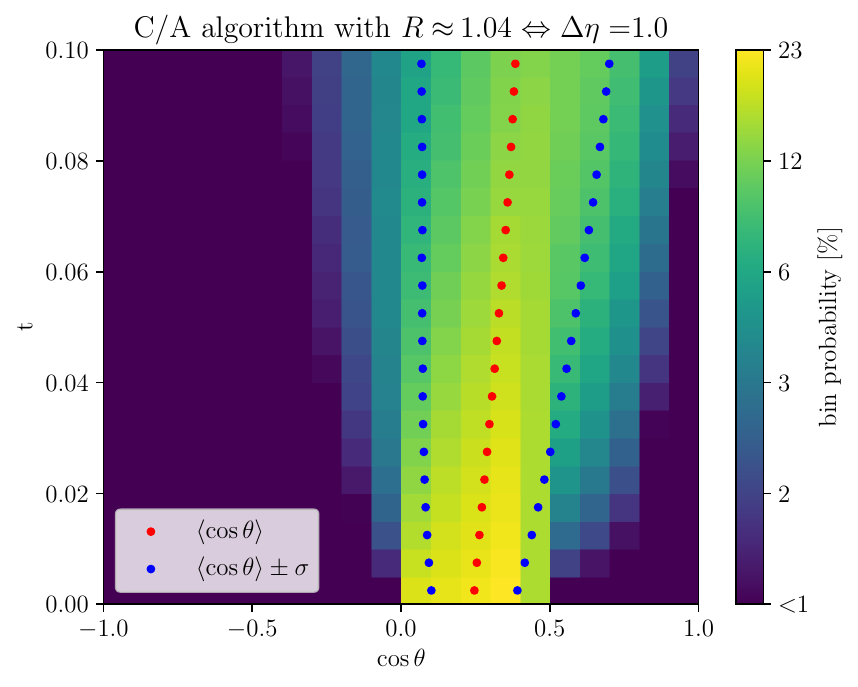}
    \end{subfigure}
    \hfill
    \begin{subfigure}{0.32\linewidth}
        \includegraphics[width=\linewidth]{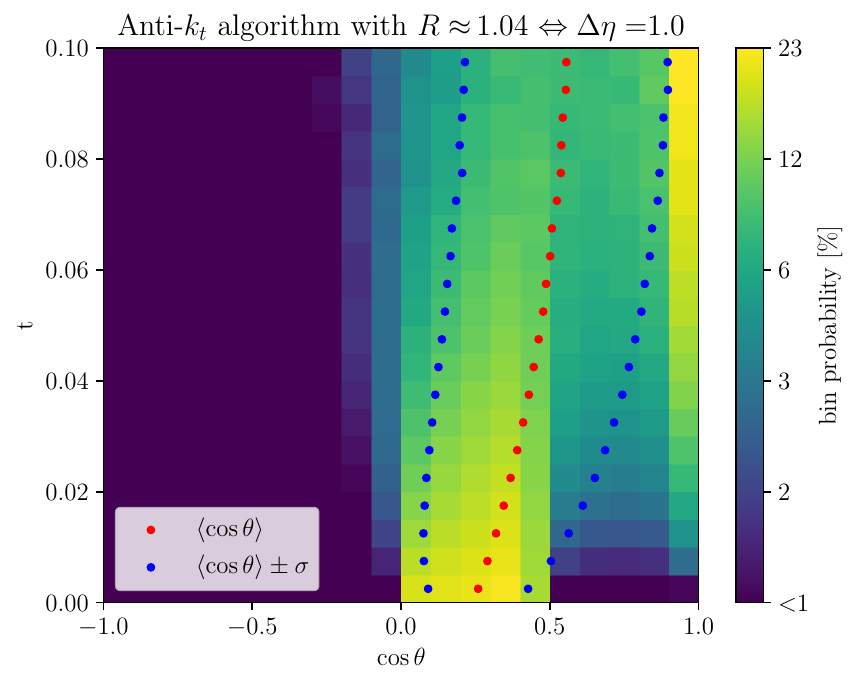}
    \end{subfigure}
    
    \medskip
    
    \begin{subfigure}{0.32\linewidth}
        \includegraphics[width=\linewidth]{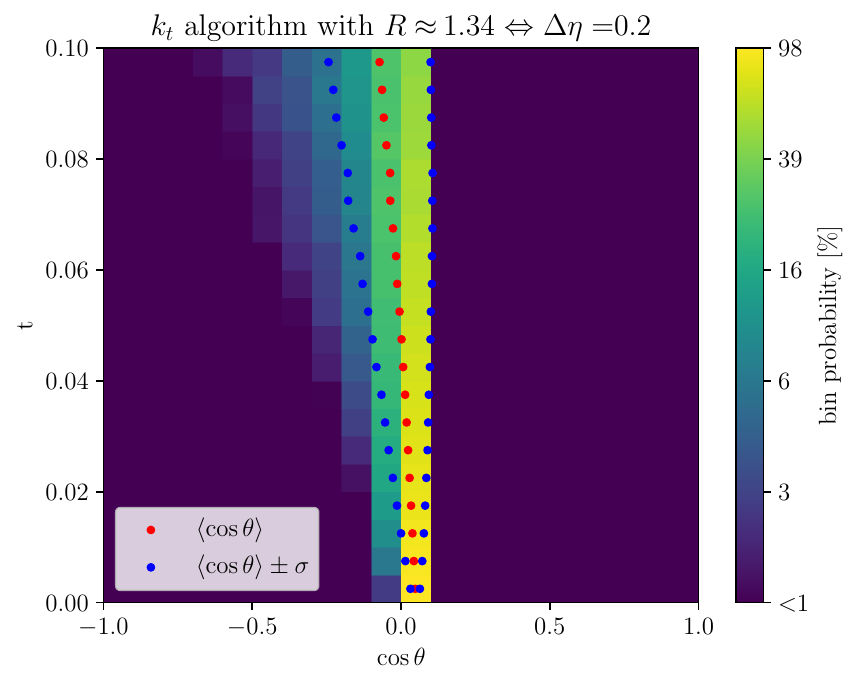}
    \end{subfigure}
    \hfill
    \begin{subfigure}{0.32\linewidth}
        \includegraphics[width=\linewidth]{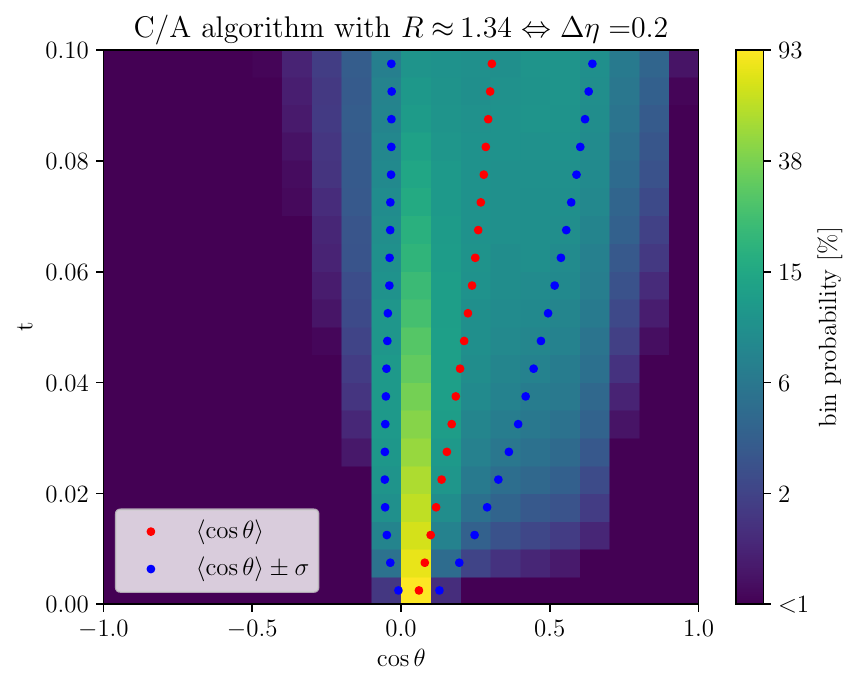}
    \end{subfigure}
    \hfill
    \begin{subfigure}{0.32\linewidth}
        \includegraphics[width=\linewidth]{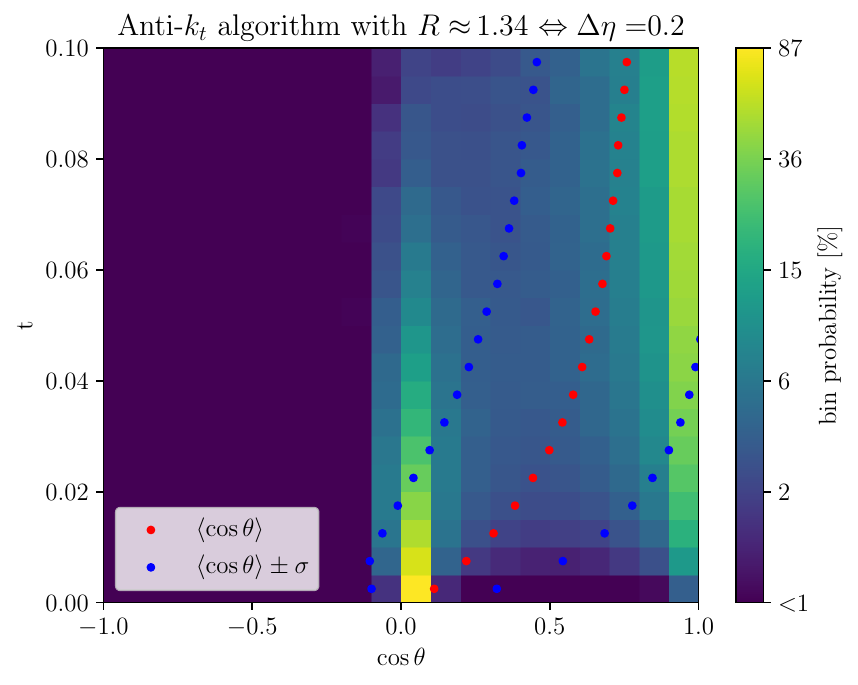}
    \end{subfigure}
    
    \caption{Probability distributions for the cosine of the smaller emission angle, for emissions into the out-region $A_{\mathrm{gap}}$ at shower time $t$. The individual plots correspond to different choices of the clustering algorithm and clustering distance $R$.}
    \label{fig:emission_plot}
\end{figure}

In the rightmost column of Figure~\ref{fig:emission_plot} we visualized this effect. The three subplots correspond to different gap sizes $\Delta\eta$. A subplot reads as follows: Horizontal slices correspond to different shower times $t$. The color represents the probability, that if a gluon $n_k$ is emitted into the gap at time $t$ from a dipole $(n_i,n_j)$, the cosine of the smaller emission angle $\cos \theta = \max(\cos\theta_{ik},\cos\theta_{jk})$ falls into one of 20 equal sized bins of $\cos \theta$ from $-1$ to $1$. We also show the mean and standard deviation of the emission angle. For example, the plot in the bottom right-hand corner is to be interpreted as follows. If radiation falls into the gap at $t=0$, then this needs to happen at an angle close to $90^\circ$, because the gap is a slim band around the equator of the sphere and the radiators $n_1$ and $n_2$ are located at the North and South Pole. With increasing shower times, the emissions occur at more and more acute angles. At shower times $t\sim 0.1$, most emissions occur in the last bin, which corresponds to emission angles smaller than $\sim 26^\circ$. This shows again, that at large $t$, the Sudakov effects are driven by collinear emissions. In general, we see for all choices of the gap size $\Delta \eta$ that the probability distribution peaks further and further to the right with increasing $t$. All in all, this explains why including only primary radiation is particularly bad in this case -- we cannot generate collinear emissions into the gap and thus one misses the most important contribution.

The same figure provides an understanding of the suppression of non-global effects for clustering algorithms. In the first and second columns of Figure~$\ref{fig:emission_plot}$, we produced angle plots for the $k_t$ and C/A algorithm. The plots for the $k_t$ algorithm are particularly simple to interpret. Here, an out-emission $n_k$ satisfies $1-R^2/2>\cos(\theta_{ik})$ for all emitters $n_i$, thus we see a sharp boundary in the color plots. This means, that the probability distribution cannot move to the right and, in particular, collinear emissions into the gap are never possible, which explains the strong suppression of non-global effects. The physics of the C/A algorithm lies somewhere in between the anti-$k_t$ and the $k_t$ algorithm. Here there is no strict cutoff on angles, in fact, in principle arbitrarily collinear emissions into the gap are possible, but strongly suppressed. This also gives a possible explanation for why we did not find any evidence for a plateauing of the gap fraction for the C/A algorithm. The suppression of Sudakov effects due to the shrinking of the effective gap is in conflict with an enhancement of Sudakov effects due to collinear radiation into the gap. We were not able to conclusively determine which effect wins out in the asymptotic limit.

We have understood why secondary-emission effects are strongly suppressed when we apply clustering algorithms to the final state. However, it might still be surprising that one ends up so close to the full result by including only primary emissions. For $k_t$ clustering with large $R$, i.e., the $\Delta\eta=0.2$ case, this might be intuitive. Here, the effective gap size tends to $0$ already for very small shower times. There is basically no time for the shower to generate emissions that are far different in direction from the primary emitters and after the gap is gone, the evolution is unitary regardless of whether we include secondary emissions or not. However, for the $k_t$ algorithm the primary approximation works nicely even for the relatively large gap sizes $\Delta\eta=1$ and $\Delta\eta=2$. The deviation due to the primary approximation is smaller than 10\%, which is comparable in size to what one expects from finite $N_c$ or NLL corrections. A hint towards a possible explanation why the approximation works so well can be found in Table~\ref{tab:primary_approximation}. For $\Delta\eta =1.0$ and $\Delta \eta =0.2$ the primary approximation actually underestimates the gap fraction, which is surprising at first glance. An explanation of the underlying mechanism can be found in Figure~\ref{fig:emission_plot}. The mean value for the emission angle into the gap actually increases with $t$ for the $k_t$ algorithm with small $\Delta\eta$, i.e., the red line bends to the left. Our interpretation is, that with secondary emissions, we can actually also generate dipoles from which it is less likely to emit into the gap than for the original primary emitters. The primary approximation seems to work so well because in the full result, secondary dipoles with higher and lower gap emission probability than the original primary dipole balance out.

\section{Summary and outlook\label{sec:conclusion}}

In this paper, we extended the class of factorization theorems for non-global observables from fixed angular constraints to observables defined using sequential jet clustering. As in the simpler case of fixed angular constraints, the associated cross sections factorize into hard and soft functions. Through the clustering, the angular constraints depend on the energy fractions of the hard partons, which requires the hard functions to be differential in the energy fractions.

In our factorization theorem based on SCET, the resummation is obtained by computing the RG evolution of the hard functions from the scale of the hard scattering down to the veto scale. The RG evolution simplifies considerably at LL accuracy because the one-loop anomalous dimension enforces strong energy ordering, which renders the energy fraction integrals trivial. We have described in detail how $k_t$-type clustering algorithms simplify in the strongly ordered limit and how the ordered clustering can be implemented efficiently in the parton shower code used to perform the resummation of the leading logarithms.

Using a shower code with energy-ordered clustering, we have computed the leading logarithms in dijet production in $e^+e^-$ collisions with a veto on the energy of emissions in the gap outside the jets. We have highlighted and explained the two main effects of the jet clustering. The first one is a shrinking of the effective gap area with increasing shower time or, equivalently, lower veto energy. The effective area is reduced because emissions are clustered out of the gap into the jets.  This clustering suppresses the Sudakov effects, which decrease the cross section for small values of the veto energy since the logarithmic corrections are proportional to the effective gap area. In extreme cases, all additional emissions get clustered into the hard jets so that the cross section becomes independent of the veto energy below a certain threshold. The second effect is a suppression of secondary radiation due to the clustering of near-collinear radiation. Both of these effects are only present for C/A and $k_t$ clustering, while the anti-$k_t$ jets retain a cone boundary since emissions are successively clustered into the primary partons. At LL accuracy, anti-$k_t$ jets are thus equivalent to cone jets, as long as the gap is defined as the region outside the jets \cite{Cacciari:2008gn}. For the $k_t$ algorithm our general findings are in line with earlier studies of the leading logarithms associated with jet clustering  \cite{Appleby:2002ke,Banfi:2005gj,Delenda:2006nf}, but we believe that the detailed analysis in Section~\ref{sec:gapsize} sheds additional light on the underlying physics. To our knowledge, we are the first to analyze clustering effects for gaps between C/A jets.

The factorization framework presented here can be used to resum subleading logarithms. For fixed angular constraints, numerical results at subleading logarithmic accuracy were presented recently \cite{Becher:2023vrh}, based on the computation of the two-loop anomalous dimension in \cite{Becher:2021urs}. The only additional ingredient needed to apply the framework to $k_t$-type jets is the extraction of the double-emission piece of the anomalous dimension, differential in the energies of the two emissions. Together with subleading logarithmic effects, finite-$N_c$ effects should be studied. The one-loop anomalous dimension we presented contains the full color information, but implementing it into a Monte Carlo framework is challenging. A simpler alternative, which should be sufficient for many cases, will be to compute the first few emissions with full color information. In addition to going to higher precision, it will be interesting to analyze more complicated observables, in particular double-logarithmic quantities such as the jet mass \cite{Kelley:2012kj,Kelley:2012zs,Delenda:2012mm,Ziani:2021dxr}, or more generally jet substructure observables, see e.g.\ \cite{Dasgupta:2013ihk,Larkoski:2017jix}.

\begin{acknowledgments}     
We thank Nicolas Schalch for comments on the manuscript. This work grew out of a one-semester visit of JH at the University of Bern. We thank Massimiliano Grazzini for arranging and supporting this visit.
The work of JH is supported by the Swiss National Science Foundation (SNSF) under contract 200020\_188464.
\end{acknowledgments}

\end{document}